\documentclass[a4paper,11pt]{article}

\usepackage{amsmath,amssymb,bbm}
\usepackage{color}
\usepackage{graphicx}
\usepackage{epstopdf}
\usepackage[normalem]{ulem}
\usepackage[bookmarks=true,bookmarksnumbered=true,bookmarkstype=toc]{hyperref}
\hypersetup{
pdftitle={World-volume Effective Actions of Exotic Five-branes},
pdfauthor={Tetsuji Kimura, Shin Sasaki and Masaya Yata},
colorlinks={true},
linkcolor={blue},
citecolor={blue},
urlcolor={blue},
filecolor={black}
}
\makeatletter

 \@addtoreset{equation}{section}
\makeatother

\newcounter{Enumerate}

\DeclareFontFamily{U}{rsf}{}
\DeclareFontShape{U}{rsf}{m}{n}{
  <5> <6> rsfs5 <7> <8> <9> rsfs7 <10-> rsfs10}{}
\DeclareMathAlphabet\Scr{U}{rsf}{m}{n}

\usepackage[mathscr]{eucal}

\newcommand{\del}{\partial}

\newcommand{\LS}{\ \ \ \ \ \ \ \ \ \ }
\newcommand{\ls}{\ \ \ \ \ }
\newcommand{\wt}{\widetilde}
\newcommand{\wh}{\widehat}
\newcommand{\ve}{\varepsilon}

\newcommand{\bsubeq}{\begin{subequations}}
\newcommand{\esubeq}{\end{subequations}}
\newcommand{\noi}{\noindent}

\newcommand{\nn}{\nonumber}

\renewcommand{\d}{{\rm d}}
\newcommand{\e}{{\rm e}}
\renewcommand{\i}{{\rm i}}
\renewcommand{\l}{\ell}
\newcommand{\w}{\wedge}

\newcommand{\slb}{\scalebox}

\renewcommand{\Im}{{\rm Im}}
\renewcommand{\Re}{{\rm Re}}

\def\+{{+\!\!\!+}} 

\begin{document}
\allowdisplaybreaks{

\thispagestyle{empty}

\begin{flushright}
KEK-TH-1725
\\
TIT/HEP-635
\end{flushright}

\vspace{25mm}

\noi
\slb{2.5}{World-volume Effective Actions of}

\vspace{5mm}

\noi
\slb{2.5}{Exotic Five-branes}

\vspace{15mm}

\slb{1.2}{Tetsuji {\sc Kimura}$^{\,a}$}, \
\slb{1.2}{Shin {\sc Sasaki}$^{\,b}$} 
\ and \ 
\slb{1.2}{Masaya {\sc Yata}$^{\,c}$}

\slb{.85}{\renewcommand{\arraystretch}{1.2}
\begin{tabular}{rl}
$a$ & 
{\sl 
Department of Physics,
Tokyo Institute of Technology} 
\\
& {\sl 
Tokyo 152-8551, JAPAN
}
\\
& {\tt tetsuji \_at\_ th.phys.titech.ac.jp}
\end{tabular}
}

\slb{.85}{\renewcommand{\arraystretch}{1.2}
\begin{tabular}{rl}
$b$ & {\sl
Department of Physics,
Kitasato University}
\\
& {\sl 
Sagamihara 252-0373, JAPAN}
\\
& {\tt shin-s \_at\_ kitasato-u.ac.jp}
\end{tabular}
}

\slb{.85}{\renewcommand{\arraystretch}{1.2}
\begin{tabular}{rl}
$c$ & {\sl
KEK Theory Center, Institute of Particle and Nuclear Studies,}
\\
& {\sl 
High Energy Accelerator Research Organization (KEK)
}
\\
& {\sl 
Tsukuba, Ibaraki 305-0801, JAPAN}
\\
& {\tt yata \_at\_ post.kek.jp}
\end{tabular}
}

\vspace{10mm}


\noindent
\slb{1.1}{\sc Abstract}:
\begin{center}
\slb{.95}{
\begin{minipage}{.95\textwidth}
We construct world-volume effective actions of exotic $5^2_2$-branes in
 type IIA and IIB string theories. 
The effective actions are given in fully space-time covariant forms with two
 Killing vectors associated with background isometries.
The effective theories are governed by the six-dimensional 
$\mathcal{N} = (2,0)$ tensor multiplet and 
$\mathcal{N} = (1,1)$ vector multiplet, respectively.
Performing the S-duality transformation to the $5^2_2$-brane effective
 action in type IIB string theory,
we also work out the world-volume action of the $5^2_3$-brane.
We discuss some additional issues relevant to the exotic five-branes in
 type I and heterotic string theories.
\end{minipage}
}
\end{center}

\newpage
\section{Introduction}
D-branes are key ingredients to understand the whole picture of
string theories. 
It is known that the D-branes play a part in non-perturbative
effects of string and gauge theories and have been studied extensively.
D-branes are dynamical objects and have its
geometrical origin in supergravity theories.
There are other various extended objects in string theories, such as Kaluza-Klein (KK) monopoles and NS5-branes.
They are also geometrical objects in supergravity theories.
The meaning of ``geometrical'' is that corresponding solutions in supergravity are 
single-valued in space-time.
The D-branes, KK-monopoles and NS5-branes are sources of the R-R, KK vector and NS-NS fluxes.
Physical properties of these objects have been studied from various
viewpoints.

On the other hand, other types of less known extended objects, which are
called {\it exotic branes} (also known as Q-, and R-branes) 
\cite{Eyras:1999at, deBoer:2010ud, Hassler:2013wsa}, are beginning to attract
attention. 
The exotic branes are obtained by U-duality transformations of the
geometric branes \cite{Elitzur:1997zn, Obers:1998fb}. 
It has been pointed out that the exotic branes play an important role in
the study of the blackhole microstates and the polarization of D-branes 
\cite{deBoer:2010ud, Kikuchi:2012za, deBoer:2012ma}. 
Similar to the fact that the geometric branes are sources of the ordinary
supergravity fluxes, the exotic branes are sources of {\it non-geometric fluxes} \cite{Hull:2004in, Hassler:2013wsa}. 
The non-geometric fluxes are discussed extensively in the context of
flux compactifications \cite{Dabholkar:2002sy, Kachru:2002sk}.
The group theoretical classification of the exotic branes have been
studied in the past.
However, 
only the limited physical properties of the exotic
branes are known.
There are studies on the exotic branes from the framework
of supergravity 
\cite{Elitzur:1997zn, Obers:1998fb, LozanoTellechea:2000mc, deBoer:2010ud, Bergshoeff:2011zk, Bergshoeff:2011se, Bergshoeff:2012pm, deBoer:2012ma}, 
string world-sheet theories 
\cite{Kimura:2013fda, Kimura:2013zva, Kimura:2013khz}, 
double field theory \cite{Hull:2006va, Albertsson:2008gq, Hull:2009mi, 
Albertsson:2011ux, Hohm:2013bwa, Andriot:2014uda, Berkeley:2014nza} and so on.
Among other things, the world-volume effective theory
\cite{Eyras:1999at, Chatzistavrakidis:2013jqa} 
is the most direct approach to study the dynamics of the exotic branes.

The purpose of this paper is to construct world-volume effective actions
of the exotic five-branes in type IIA and IIB string theories.
We focus on the exotic $5^2_2$-branes and $5^2_3$-branes that are obtained from the
NS5-branes via T- and S-dualities \cite{deBoer:2012ma}.
Compared with the preceding work \cite{Chatzistavrakidis:2013jqa}, 
where the effective actions of the type IIB $5^2_2$-brane and the $5^2_3$-brane are studied, 
the actions in this paper will be written in the fully space-time
covariant forms.
In the actions, two Killing vectors associated with the two isometries of the backgrounds
are manifest. We also explicitly write down the effective action
of the type IIA $5^2_2$-brane in a manifestly Lorentz invariant manner.
After the gauge fixing, the massless fields of the
world-volume theories 
are organized into the 
$\mathcal{N} = (2,0)$ tensor multiplet (IIA) and the
$\mathcal{N} = (1,1)$ vector multiplet (IIB)
in six dimensions. 
In the pioneered work \cite{Eyras:1999at}, 
the world-volume effective actions of the exotic $6^1_3$-brane and
seven-branes are studied where the geometric part of the actions are written
as gauged sigma models when the NS-NS $B$-field is turned off.
A specific property of the effective actions of the five-branes in this
paper is its dependence on the two Killing vectors $k^{\mu}_1$,
$k^{\mu}_2$. We will see that the actions are not written as gauged
sigma models when all the closed string backgrounds are involved.

We also discuss some properties of exotic five-branes in type I and heterotic string theories. 
The orientifold projection of type IIB theory and duality chains
indicate that exotic five-branes appear also in type I and heterotic theories.
To make contact with the exotic branes in these theories, we
analyze the effective actions of the $5^2_3$-brane and $5^2_2$-brane in type
I and heterotic theories.

The organization of this paper is as follows. 
In Section \ref{sect-review522},
we review the basic properties of the exotic
$5^2_2$-brane in string theories.
We also demonstrate the string duality chains on various five-branes.
In Section \ref{sect-WVII}, we explicitly write down the world-volume effective
actions of the exotic $5^2_2$-brane and $5^2_3$-brane in IIB string theory.
We also find the effective action of the $5^2_2$-brane in type IIA theory.
The gauge symmetries of the effective theories are discussed.
In Section \ref{sect-WVI}, we discuss exotic five-branes in type I and
$SO(32)$, $E_8 \times E_8$ heterotic string
theories. The Abelian part of the effective actions of the
$5^2_3$-brane in type I and the $5^2_2$-brane in $SO(32)$, $E_8 \times E_8$ heterotic string
theories are proposed. 
Section \ref{sect-conclusion} is devoted to the conclusion and discussions.
Conventions and notations of the supergravity theories in this
paper are found in Appendix \ref{app-convention}.
In Appendix \ref{app-covBuscher}, 
the covariant T-duality transformation rules for the
NS-NS and R-R sectors are summarized. 
The explicit representations of the second T-duality transformation of
the NS-NS $B$-field and R-R potentials are given also in Appendix \ref{app-covBuscher}.

\section{Exotic five-branes and string duality chains}
\label{sect-review522}

In this section we exhibit the exotic five-branes
in string theories.
In the first subsection, we briefly review the configuration of the single $5^2_2$-brane.
This is obtained via the T-duality transformations of the configurations
of NS5-branes.
In the second subsection, we demonstrate the string duality chains on
various five-branes.

\subsection{Configuration of exotic $5^2_2$-brane}

We begin with multi-centered H-monopoles in ten-dimensional supergravity theories.
This is a smeared solution of multi-centered NS5-branes along the $x^9$-direction.
The configuration of the multi-centered H-monopoles is\footnote{We follow the conventions employed in \cite{Gregory:1997te}.} \footnote{The H-monopole geometry presented in this section corresponds to the neutral solution in
heterotic supergravity.
There are other five-brane solutions known as the gauge and the symmetric
solutions \cite{Callan:1991ky}.}
\begin{eqnarray} \label{NS5-R3S1}
\begin{aligned}
\d s^2_{\text{H}} \ &= \ 
\d x_{012345}^2
+ H (\vec{r}) \big( (\d \vec{r})^2
+ (\d x^9)^2 \big) 
\, , \ls
\vec{r} \in {\mathbb R}_{678}^3
\, , \ls
x^9 \ = \ R_9 \, \vartheta
\, , \\
H (\vec{r}) \ &= \ 
1 + \sum_p H_p
\, , \ls
H_p \ = \ 
\frac{\alpha'}{2 R_9 | \vec{r} - \vec{r}_p|} 
\, , \\
\e^{2 \phi} \ &= \ 
H (\vec{r})
\, , \ls
H_{mnp} \ = \ 
\ve_{mnp}{}^{q} \, \del_{q} \log H (\vec{r})
\, .
\end{aligned}
\end{eqnarray}
Here the H-monopoles are expanded along the $012345$-directions.
The vector $\vec{r}_p$ denotes the position of the $p$-th H-monopole in the $678$-directions.
The space-time metric $g_{\mu \nu}$, the dilaton $\phi$ and the field strength $H_{mnp}$ of the NS-NS $B$-field are given by a harmonic function $H (\vec{r})$.
The indices $m,n,p,q$ run from $6$ to $9$.
Since the harmonic function does not depend on $x^9$,
there is an isometry along this direction.
We assume that the $x^9$-direction is compactified on a circle of radius $R_9$.

We perform a T-duality transformation along the $x^9$-direction.
Following the Buscher rule \cite{Buscher:1987sk}, 
we transform the above description to
\begin{eqnarray} \label{KKM-metric}
\begin{aligned}
\d s^2_{\text{KKM}} \ &= \ 
\d x_{012345}^2
+ H (\vec{r}) \, \d x_{678}^2
+ \frac{1}{H (\vec{r})} \big( \d \wt{x}^9 + {\omega} \big)^2
\, , \ls
\vec{r} \in {\mathbb R}_{678}^3
\, , \\
H (\vec{r}) \ &= \ 
1 + \sum_p H_p
\, , \ls
H_p \ = \ 
\frac{\wt{R}_9}{2| \vec{r} - \vec{r}_p|}
\, , \ls
\wt{x}^9 \ = \ \wt{R}_9 \, \wt{\vartheta}
\, , \ls
\wt{R}_9 \ = \ \frac{\alpha'}{R_9}
\, , \\
\d {\omega} \ &= \ 
*_3 \d H
\, , \ls
\e^{2 \phi} \ = \ 1
\, , \ls
B \ = \ 
\beta \, \d \Big[ \frac{1}{H} \Big( \d \wt{x}^9 + {\omega} \Big)
\Big]
\, ,
\end{aligned}
\end{eqnarray}
where $\beta$ is a constant.
This is the configuration of multi-centered KK-monopoles (KK5-branes).
The $9$-th coordinate $x^9$ is transformed to $\wt{x}^9$.
This direction is a compact circle of radius $\wt{R}_9$.
Here the dilaton and the NS-NS $B$-field are trivial up to total derivatives.
Instead, the KK-vector $\omega$ is involved in the space-time metric.
The relation between the KK-vector $\omega$ and the harmonic function $H(\vec{r})$ is given as a monopole equation.
Indeed the transverse space of the multi-centered KK-monopoles is given as a multi-centered Taub-NUT space.

There are no more directions with isometry in the configuration of the KK-monopoles (\ref{KKM-metric}).
However, if an infinite number of KK-monopoles are arrayed along the $x^8$-th direction, 
we can see a shift symmetry along it.
Under this setup, the configuration (\ref{KKM-metric}) has an additional isometry along the $x^8$-direction.
Compactifying the $x^8$-direction with radius $R_8$
and performing a T-duality transformation along it, we obtain 
the dual coordinate $\wt{x}^8$ of radius $\wt{R}_8 = \alpha'/R_8$ and
the following configuration:
\begin{eqnarray} \label{single522}
\begin{aligned}
\d s_{5_2^2}^2 \ &= \ 
\d x_{012345}^2
+ H \, \big( \d \varrho^2 + \varrho^2 (\d \vartheta_{\!\varrho})^2 \big)
+ \frac{H}{K} \big( (\d \wt{x}^8)^2 + (\d \wt{x}^9)^2 \big)
\, , \\
\e^{2 \phi} \ &= \ \frac{H}{K}
\, , \ls
B \ = \ 
- \frac{\sigma'' \vartheta_{\!\varrho}}{K} \, \d \wt{x}^8 \w \d \wt{x}^9
\, , \ls
B^{(8,2)} \ = \ 
- \frac{K}{H} \, \d x^0 \w \d x^1 \w \cdots \w \d x^5
\, , \\
H \ &= \ 
\sigma'' \, \log \frac{\mu''}{\varrho}
\ = \ 
h_0 + \sigma'' \, \log \frac{\mu_0}{\varrho}
\, , \ls
K \ = \ H^2 + (\sigma'' \vartheta_{\!\varrho})^2
\, , \ls
\sigma'' \ = \ 
\frac{\wt{R}_8 \wt{R}_9}{2 \pi \alpha'}
\, .
\end{aligned}
\end{eqnarray}
This is the background geometry of the exotic $5^2_2$-brane expanded along the $012345$-directions.
The transverse space is locally described as ${\mathbb R}^2 \times T^2$ whose coordinates are expressed in terms of $\{ \varrho, \vartheta_{\varrho}; \wt{x}^8, \wt{x}^9 \}$.
$B^{(8,2)}$ is a six-form whose transverse space has two isometries.
We use the notation $B^{(8,2)}$ instead of $B^{(6)}$ to denote the Hodge
dual of the T-dualized $B$-field.
This is natural since the $5^2_2$-brane couples to $B^{(6)}$ which
is T-duality transformed along two isometries.
$B^{(8,2)}$ is also called the mixed-symmetry tensor \cite{Meessen:1998qm, Bergshoeff:2011zk}.
In this configuration, 
the space-time metric, the dilaton, and the NS-NS $B$-field are described by not only the harmonic function $H$ but also the angular coordinate $\vartheta_{\rho}$.
The harmonic function diverges in the IR region.
This implies that asymptotic description of the single $5^2_2$-brane is not well-defined.
Furthermore, caused by the dependence of the angular coordinate,
the space-time metric, the dilaton field, and the NS-NS $B$-field are no
longer single-valued. 
Due to the multi-valuedness of the solution, there is a non-trivial
monodromy around the $5^2_2$-brane.
The monodromy characterizes a kind of conserved charge (Page charge) of the brane \cite{deBoer:2012ma}.

\subsection{String duality chains on five-branes}

In this subsection we demonstrate the string duality chains on various five-branes. 
We begin with the following M-theory branes:
an M5-brane, a KK6-brane, and a $5^3$-brane.
They are related to five-branes in string theories.
The former two are standard branes, whilst the latter one is an exotic object \cite{deBoer:2010ud}.
Their tensions are evaluated in terms of the Planck length $\l_{\text{p}}$ in eleven dimensions and radii of compact circles:
\begin{eqnarray} \label{branesM}
\begin{aligned}
\text{M5(12345)} \ &: &\ \ 
M \ &= \ 
\frac{1}{\l_{\text{p}}^6}
\, , \\
\text{$5^3$(12345,89$\natural$)} \ &: &\ \ 
M \ &= \ 
\frac{(R_8 R_9 R_{\natural})^2}{\l_{\text{p}}^{12}}
\, , \\
\text{KK6(12345$\natural$,9)} \ &: &\ \ 
M \ &= \ 
\frac{(R_9)^2}{\l_{\text{p}}^9}
\, .
\end{aligned}
\end{eqnarray}
We followed the conventions for branes employed in \cite{deBoer:2012ma}.
The parameter $R_{\natural}$ in the mass of the $5^3$-brane is the radius of M-theory circle.
The transverse directions of the M5-brane is a topologically flat five-dimensional space, while the transverse space of the $5^3$-brane has three isometry directions.
The transverse four-directions of the KK6-brane is the Taub-NUT space.
All of the above branes are coupled to the six-form potentials, the dual of the three-form potential, in M-theory.

In type IIA string theory,
we focus on the following five-branes and six-branes\footnote{Here we do
not consider the D6-brane because this is not directly related to exotic
five-branes in this work.}:
\begin{eqnarray} \label{branesIIA}
\begin{aligned}
\text{NS5(12345)} \ &: &\ \ 
M \ &= \ 
\frac{1}{g_{\text{st}}^2 \l_{\text{st}}^6}
\, , \\
\text{KK5(12345,9)} \ &: &\ \ 
M \ &= \ 
\frac{(R_9)^2}{g_{\text{st}}^2 \l_{\text{st}}^{8}}
\, , \\
\text{$5^2_2$(12345,89)} \ &: &\ \ 
M \ &= \ 
\frac{(R_8 R_9)^2}{g_{\text{st}}^{2} \l_{\text{st}}^{10}}
\, , \\
\text{$6^1_3$(123458,9)} \ &: &\ \ 
M \ &= \ 
\frac{(R_9)^2}{g_{\text{st}}^3 \l_{\text{st}}^{11}}
\, .
\end{aligned}
\end{eqnarray}
Here the eleven-dimensional Planck length $\l_{\text{p}}$ is described by the string coupling constant $g_{\text{st}}$ and the string length $\l_{\text{st}} = \sqrt{\alpha'}$ in such a way as 
$\l_{\text{p}} = g_{\text{st}} \l_{\text{st}}^3$, 
where $\alpha'$ is the Regge-slope parameter.
The parameter $R_9$ in the mass of the KK5-brane is the radius of the Taub-NUT circle.
The radius of the M-theory circle $R_{\natural}$ is also given as
$R_{\natural} = g_{\text{st}} \l_{\text{st}}$.
The NS5-brane and the KK5-brane are obtained by the dimensional
reduction of one direction transverse to the M5-brane and longitudinal to 
the KK6-brane, respectively.
We find the $5^2_2$-brane by the dimensional reduction of the $5^3$-brane 
along the direction of the M-theory circle.
The $6^1_3$-brane can be found by the dimensional reduction 
of the KK6-brane along a transverse direction in ${\mathbb R}^3$ of the
Taub-NUT space.
We remark that the NS5-brane, the KK5-brane, and the $5^2_2$-brane are
coupled to the NS-NS six-form potentials, which are dual to the NS-NS two-form potential,
while the $6^1_3$-brane is coupled to a R-R seven-form potential as a mixed-symmetry tensor.

We move to type IIB string theory from type IIA string theory.
The above branes (\ref{branesIIA}) in type IIA theory are 
transformed to the other branes under the following string duality transformation rules:
\begin{eqnarray}
\begin{aligned}
\text{T}_i \ &: &\ \ 
R_i \ &\to \ \frac{\l_{\text{st}}^2}{R_i}  
\, , &\ls
g_{\text{st}} \ &\to \ \frac{\l_{\text{st}}}{R_i} \, g_{\text{st}}
\, , \\
\text{S} \ &: &\ \ 
g_{\text{st}}^{\vphantom{-1}} \ &\to \ g_{\text{st}}^{-1}
\, , &\ls
\l_{\text{st}} \ &\to \ g_{\text{st}}^{1/2} \l_{\text{st}}^{\vphantom{-1}}
\, .
\end{aligned}
\end{eqnarray}
Here $\text{T}_i$ denotes the T-duality transformation along the $i$-th
direction and S is the S-duality transformation.
The list of five-branes in type IIB string theory is as follows:
\begin{eqnarray}
\begin{aligned}
\text{D5(12345)} \ &: &\ \ 
M \ &= \ 
\frac{1}{g_{\text{st}} \l_{\text{st}}^6}
\, , \\
\text{NS5(12345)} \ &: &\ \ 
M \ &= \ 
\frac{1}{g_{\text{st}}^2 \l_{\text{st}}^6}
\, , \\
\text{KK5(12345,9)} \ &: &\ \ 
M \ &= \ 
\frac{(R_9)^2}{g_{\text{st}}^2 \l_{\text{st}}^{8}}
\, , \\
\text{$5^2_2$(12345,89)} \ &: &\ \ 
M \ &= \ 
\frac{(R_8 R_9)^2}{g_{\text{st}}^{2} \l_{\text{st}}^{10}}
\, , \\
\text{$5^2_3$(12345,89)} \ &: &\ \ 
M \ &= \ 
\frac{(R_8 R_9)^2}{g_{\text{st}}^3 \l_{\text{st}}^{10}}
\, .
\end{aligned}
\end{eqnarray}
The NS5-brane in type IIB theory is obtained via the T-duality transformation along the longitudinal (transverse) direction of the NS5-brane (the KK5-brane) in type IIA theory.
The KK5-brane and the $5^2_2$-brane are also obtained via the T-duality transformations along suitable directions of five-branes in type IIA theory.
The D5-brane and the $5^2_3$-brane are obtained via the S-duality transformation of the NS5-brane and the $5^2_2$-brane in type IIB theory, respectively.
We remark that there are other five-branes in type II string theories.
We obtain five-branes of co-dimensions less than two by 
performing further duality transformations.
For example, the $5^{(1,2)}_3$-brane in type IIA theory is obtained via
the T-duality transformation along the 7-th direction of the
$5^2_3$-brane in type IIB theory. A list of five-branes in type II
string theories is found in Figure \ref{fig1}.
In the following, our main focus is on the five-branes which have co-dimension
two or more.
\begin{figure}[t]
\begin{center}
\includegraphics[scale=0.85]{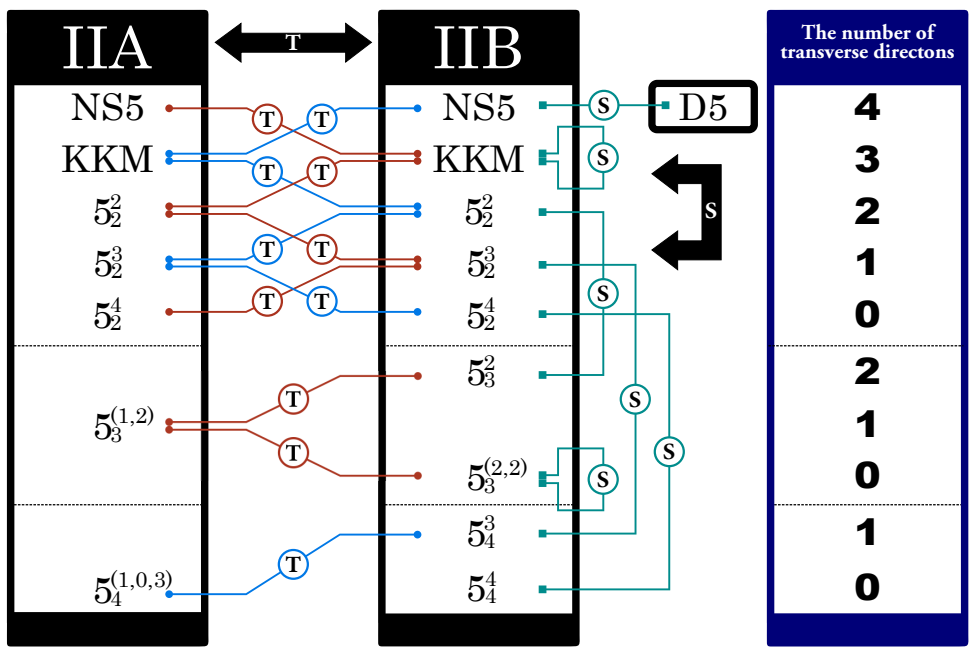}
\begin{minipage}{.95\textwidth}
\caption{\sl\small 
Type II five-branes which are obtained from the NS5-branes by duality chains.
}
\label{fig1}
\end{minipage}
\end{center}
\end{figure}

We further move to type I string theory via the orientifold projection of type IIB theory.
Due to this projection, the NS-NS $B$-field and its dual are projected out.
Then the five-branes surviving in type I theory are the D5-brane and the
$5^2_3$-brane.
Performing the S- and T-duality transformations to the five-branes in type I theory,
we can also discuss the NS5-brane, the KK5-brane, and the $5^2_2$-brane
in heterotic string theories.
They are also derived from the branes (\ref{branesM}) in M-theory via the $S^1/{\mathbb Z}_2$ compactification.
We discuss the world-volume effective actions of five-branes in type I
and heterotic string theories in Section \ref{sect-WVI}.
We summarize the above demonstration of the string duality chains on
five-branes in Figure \ref{fig2}.
\begin{figure}[t]
\begin{center}
\includegraphics[scale=0.85]{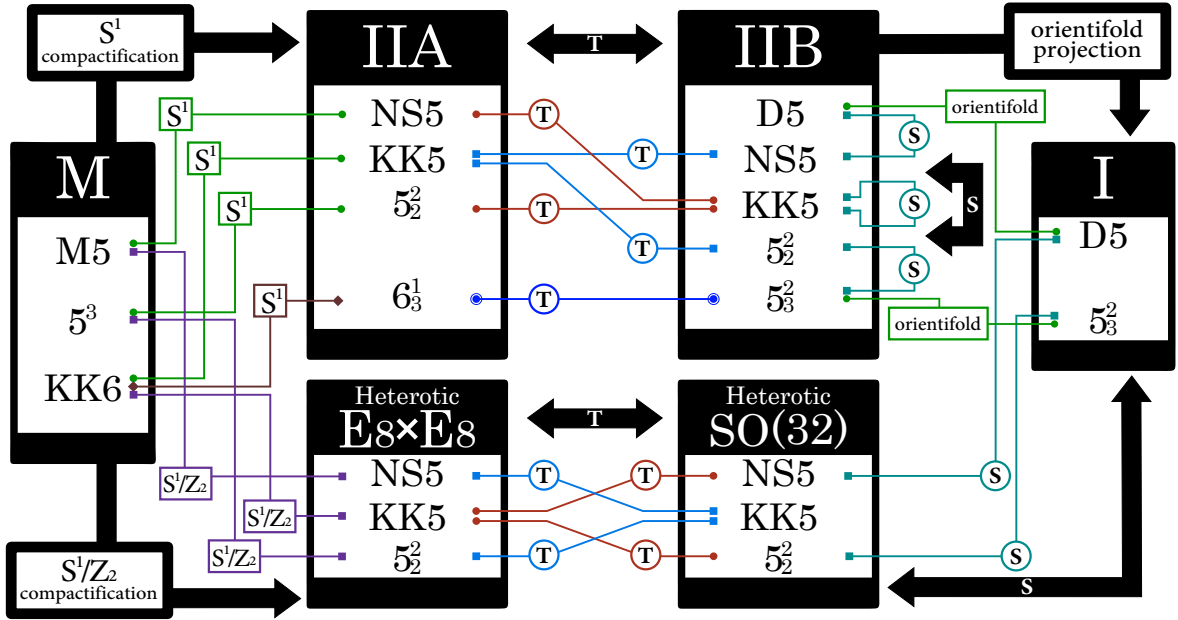}
\begin{minipage}{.95\textwidth}
\caption{\sl\small String duality chains on various five-branes.
}
\label{fig2}
\end{minipage}
\end{center}
\end{figure}

Recently, the ``exotic'' configurations have been investigated very well 
in supergravity theories \cite{deBoer:2010ud, deBoer:2012ma},
and in the framework of double field theory \cite{Kikuchi:2012za}.
In our previous works \cite{Kimura:2013fda, Kimura:2013zva, Kimura:2013khz}
we studied the worldsheet sigma model whose target space is the
background geometry of the exotic $5^2_2$-brane.
In the next section, 
we discuss the world-volume descriptions of the exotic five-branes in type IIA and IIB string theories.

\section{World-volume effective actions of type II $5^2_2$-brane and $5^2_3$-brane}
\label{sect-WVII}

In this section, we explicitly write down the bosonic part of the world-volume effective
actions of the single exotic $5^2_2$-branes in type IIA and IIB, 
and the $5^2_3$-brane in type IIB string theories. 
We consider the half-BPS five-branes which preserve sixteen supercharges in its world-volume.
The supermultiplets in the world-volume theories of the NS5-branes and
the KK5-branes in type I, II and heterotic theories are classified in
\cite{Hull:1997kt}. 
With the relations by the T-duality, 
the world-volume theories of the type IIB $5^2_2$-brane and $5^2_3$-brane are governed by the six-dimensional $\mathcal{N} =(1,1)$ vector multiplet.
On the other hand, the type IIA $5^2_2$-brane world-volume theory is described by the $\mathcal{N} = (2,0)$ tensor multiplet.
We will work out the effective actions by the duality transformations of
the geometric branes.
The transformations include those of the space-time backgrounds and the
world-volume fields.
Generically, the scalar zero-modes associated with the isometry
directions are translated into the new scalar fields which do not have the geometrical meaning (it is not the transverse fluctuation modes).
The world-volume $U(1)$ gauge field in a brane becomes another
$U(1)$ gauge field by the duality transformation.
The new gauge field has a different gauge transformation rule from the original one.
In the following subsections, we consider the type IIA and IIB theories separately.

\subsection{$5^2_2$-brane and $5^2_3$-brane in type IIB theory}

As we demonstrated in Section \ref{sect-review522}, 
the type IIB $5^2_2$-brane is obtained by the T-duality transformations along the transverse directions of the NS5-brane which is S-dual of the D5-brane. 
The effective action of the single D$p$-brane in type II theories is the sum of the
Dirac-Born-Infeld (DBI) part and the Wess-Zumino part:
\begin{align}
S_{\text{D$p$}}
\ =& \ 
- T_{\text{D$p$}} \int \d^{p+1} \xi \,
\e^{-\phi} \sqrt{- \det \big( P[g]_{ab} + P[B]_{ab} + \lambda F_{ab} \big)
}
\nn \\
\ & \ 
+ \mu_p \int_{\mathcal{M}_{p+1}} \! P[\sum_n C^{(n)} \wedge \e^{B}]
 \wedge \e^{\lambda F}
\, , \label{eq:Dp-brane_action}
\end{align}
where $g_{\mu \nu}$, $B_{\mu \nu}$, $\phi$ are the background space-time metric, the NS-NS $B$-field and the dilaton. 
The fields $C^{(n)}$ are the R-R potentials and
$F_{ab} = \partial_a A_b - \partial_b A_a$ is the field strength of the $U(1)$ gauge field $A_a$.
The tension and the R-R charge of the BPS D$p$-brane are given by 
$T_{\text{D$p$}} = \mu_p = \frac{1}{(2\pi)^p g_{\text{st}}} \alpha'{}^{- (p+1)/2}$. 
We also defined $\lambda = 2\pi \alpha'$.
The symbol $P [\Phi]$ stands for the pull-back of the space-time tensor
fields $\Phi_{\mu_1\cdots \mu_N}$ onto the world-volume:
\begin{align}
P[\Phi]_{a_1 \cdots a_N} 
\ &= \ 
\Phi_{\mu_1 \cdots \mu_N} \partial_{a_1} X^{\mu_1} \cdots  \partial_{a_N} X^{\mu_N}
\, .
\end{align}
Here $\partial_a$ is the derivative with respect to the world-volume
coordinate $\xi^a$.
The summation $\sum_n$ in the Wess-Zumino part is understood such that
the integration over the world-volume dimension is well-defined. 
For example, the Wess-Zumino term of the D5-brane is given by 
\begin{align}
\mu_5 \int_{\mathcal{M}_6} \! 
\left[
P[C^{(6)}] + P[C^{(4)}] \wedge (P[B] + \lambda F) + \frac{1}{2!} P[C^{(2)}] \wedge
 (P[B] + \lambda F)^2 
+ \frac{1}{3!} C^{(0)} (P[B] + \lambda F)^3
\right]
\, .
\end{align}
The type IIB NS5-brane is S-dual of the D5-brane. The S-duality transformation rules of the background fields \cite{Chatzistavrakidis:2013jqa} are
\begin{eqnarray}
\begin{aligned}
\tau \ &\xrightarrow[\text{S}]{} \ 
- \frac{1}{\tau}
\, , \ls 
C^{(2)} \ \xrightarrow[\text{S}]{} \ B
\, , &\ls
B \ &\xrightarrow[\text{S}]{} \ - C^{(2)}
\, , \ls 
g_{\mu \nu} 
\ \xrightarrow[\text{S}]{} \ 
|\tau| \, g_{\mu \nu}
\, , \\
C^{(4)} \ &\xrightarrow[\text{S}]{} \ 
C^{(4)} + C^{(2)} \wedge B
\, , &\ls
C^{(6)} \ &\xrightarrow[\text{S}]{} \ 
- B^{(6)} + \frac{1}{2} B \wedge C^{(2)} \wedge C^{(2)}
\, ,
\end{aligned}
\label{eq:S-duality}
\end{eqnarray}
where $\tau = C^{(0)} + \i \, \e^{-\phi}$ is the complex axio-dilaton
field and $B^{(6)}$ is the magnetic dual of $B$.
Together with the space-time fields, the world-volume gauge field $A_a$
is transformed as $A_a \xrightarrow[\text{S}]{} \mathcal{A}_a$.
Then the effective action of the type IIB NS5-brane \cite{Eyras:1998hn} is
\begin{align}
S^{\text{IIB}}_{\text{NS5}}
\ =& \ 
- T_{\text{NS5}} \int \! \d^6 \xi \, \e^{-2\phi} 
\sqrt{1 + \e^{2\phi} (C^{(0)})^2} 
\sqrt{
- \det \Big( g_{\mu \nu} \, \partial_a X^{\mu} \partial_b X^{\nu} 
+ \frac{\lambda \, \e^{\phi}}{\sqrt{1 + \e^{2\phi} (C^{(0)})^2 }} \, \mathcal{F}_{ab}
\Big)
}
\nn \\
\ & \
- \mu_5 \int_{\mathcal{M}_6} 
\left[
P[B^{(6)}] - \frac{1}{2} P[B \wedge C^{(2)} \wedge C^{(2)}] 
- \lambda P[C^{(4)} + C^{(2)} \wedge B] \wedge \mathcal{F}
\right. 
\nn \\
\ & \LS \ls \
\left.
- \frac{\lambda^2}{2} P[B] \wedge \mathcal{F} \wedge \mathcal{F}
+ \frac{\lambda^3}{3!} \frac{C^{(0)}}{(C^{(0)})^2 + \e^{-2\phi}} 
\mathcal{F} \w \mathcal{F} \w \mathcal{F} 
\right]
\, .
\label{eq:IIBNS5}
\end{align}
Here $\mathcal{F}_{ab} = \partial_a \mathcal{A}_b - \partial_b \mathcal{A}_a -
\lambda^{-1} C^{(2)}_{\mu \nu} \, \partial_a X^{\mu} \partial_b X^{\nu}$.
The tension of the NS5-brane is defined as $T_{\text{NS5}} =
g_{\text{st}}^{-1} T_{\text{D5}}$.
The action \eqref{eq:IIBNS5} is invariant under the following space-time and world-volume
gauge transformations:
\begin{eqnarray}
\begin{aligned}
\delta B \ &= \ \d \Lambda^{(1)}
\, , \ls
\delta B^{(6)} \ = \
\d \Lambda^{(5)} 
- \d \lambda^{(3)} \wedge C^{(2)} 
+ \d \lambda^{(1)} \wedge B \wedge C^{(2)}
\, , \\
\delta C^{(0)} \ &= \ 0
\, , \ls
\delta C^{(2)} \ = \ \d \lambda^{(1)}
\, , \ls
\delta C^{(4)} \ = \ \d \lambda^{(3)} - B \wedge \d \lambda^{(1)}
\, , \\
\delta \mathcal{A}_a \ &= \ \partial_a \chi + \lambda^{-1} P[\lambda^{(1)}]_a
\, , \ls
\delta \mathcal{F}_{ab} \ = \ 0
\, ,
\end{aligned}
\end{eqnarray}
where $\Lambda^{(n)}$ and $\lambda^{(n)}$ ($n = 1,3,5$) are space-time gauge
parameter $n$-forms and $\chi$ is the world-volume gauge parameter.
The effective action of the KK-monopole in type IIA string theory is
obtained by the T-duality transformation of the action \eqref{eq:IIBNS5} 
along the one transverse direction of the brane \cite{Eyras:1998hn}.

The type IIB $5^2_2$-brane effective action is obtained by 
the T-duality transformations of the action \eqref{eq:IIBNS5} along the two transverse directions. 
We consider the isometries of the background fields.
The isometries are generated by two Killing vectors 
$k^{\mu}_I$ $(I=1,2)$ transverse to the NS5-brane world-volume. 
We keep the space-time covariant expression 
using the covariant Buscher rules in Appendix \ref{app-covBuscher}.
Then we find
\begin{eqnarray}
\begin{aligned}
\e^{-2\phi} \ &\xrightarrow[k_1 k_2]{} \ 
\e^{-2\phi} (\det h_{IJ}) 
\, , \ls
g_{\mu \nu} \ \xrightarrow[k_1 k_2]{} \ 
\Pi_{\mu \nu} (k_2)
+ \frac{K^{(1)}_{\mu} K^{(1)}_{\nu}}{(k_2)^2} 
- \frac{(k_2)^2 (K^{(2)}_{\mu} K^{(2)}_{\nu} - K^{(3)}_{\mu} K^{(3)}_{\nu})}{\det h_{IJ}} 
\, , \\
B \ &\xrightarrow[k_1 k_2]{} \ 
\wt{B}
\, , \\
C^{(0)} \ &\xrightarrow[k_1 k_2]{} \ 
i_{k_1} i_{k_2} (C^{(2)} + C^{(0)} B)
\, , \ls
C^{(2)} \ \xrightarrow[k_1 k_2]{} \ 
\wt{C}^{(2)}
\, , \ls
C^{(4)} \ \xrightarrow[k_1 k_2]{} \ 
\wt{C}^{(4)}
\, , 
\end{aligned}
\end{eqnarray}
where the arrow ``$\xrightarrow[k_1 k_2]{}$'' indicates that the repeated T-duality
transformations along the directions generated by the Killing vectors
, first $k^{\mu}_1$, then $k^{\mu}_2$.
The explicit forms of $\wt{B}$, $\wt{C}^{(2)}$, $\wt{C}^{(4)}$ are given in Appendix \ref{app-covBuscher}.
Here we defined the following quantities:
\begin{eqnarray}
\begin{aligned}
h_{IJ} \ &= \ 
k_I^{\mu} k_{J}^{\nu} \, (g_{\mu \nu} + B_{\mu \nu})
\, , \ls
\Pi_{\mu \nu} (k_2) \ = \
g_{\mu \nu} - \frac{1}{(k_2)^2} (i_{k_2} g)_{\mu} (i_{k_2} g)_{\nu}
\, , \\ 
(k_I)^2 \ &= \ 
g_{\mu \nu} k_I^{\mu} k_I^{\nu} \ \ (\text{no sum over $I$})
\, , \\
K^{(1)}_{\mu} \ &= \ 
(i_{k_2} B - \lambda \, \d \varphi')_{\mu}
\, , \\
K^{(2)}_{\mu} \ &= \ 
\Big(
i_{k_1} g - \frac{k_1 \cdot k_2}{(k_2)^2} (i_{k_2} g) 
+ \frac{1}{(k_2)^2} (i_{k_1} i_{k_2} B) (i_{k_2} B - \lambda \, \d \varphi')
\Big)_{\mu}
\, , \\
K^{(3)}_{\mu} \ &= \ 
\Big(
(i_{k_1} B - \lambda \, \d \varphi) 
- \frac{k_1 \cdot k_2}{(k_2)^2} (i_{k_2} B - \lambda \, \d \varphi') 
+ \frac{1}{(k_2)^2} (i_{k_1} i_{k_2} B) i_{k_2} g
\Big)_{\mu}
\, .
\end{aligned}
\label{eq:K_def}
\end{eqnarray}
The scalar fields $\varphi$ and $\varphi'$ are associated with the  
dual coordinates under the T-duality transformations along the
isometries generated by the Killing vectors $k_1^{\mu}$ and $k_2^{\mu}$,
respectively.

A salient feature of the exotic branes is its coupling to 
non-geometric fluxes.
The T-duality group theoretical representations of the potentials that couple to the solitonic branes
(i.e. objects that have the tension proportional to $\e^{-2\phi}$) 
in ten-dimensional supergravities are investigated in \cite{Bergshoeff:2011zk}.
They are anti-symmetric tensor representations of $SO(10-d,10-d)$ where $d$ is the compactified dimensions.
The conjugacy class of the solitonic branes are studied. 
In particular, the Wess-Zumino term of the co-dimension two defect
branes are discussed in \cite{Bergshoeff:2011se}\footnote{In \cite{Bergshoeff:2011se} the exotic $5^2_2$-branes, which come from the
NS5-branes by T-dualities, are called generalized KK-monopoles.}.
The explicit top coupling form of the Wess-Zumino term in the type IIB exotic
$5^2_2$-brane is studied in \cite{Chatzistavrakidis:2013jqa}.
This is defined by the T-duality transformations of the $B^{(6)}$ field
and represented by the mixed-symmetry tensor $B^{(8,2)}$ 
\cite{Meessen:1998qm, Bergshoeff:2011zk}.
We therefore obtain\footnote{Strictly speaking,
the terminology $i_{k_1} i_{k_2} B^{(8,2)}$
as the T-dualized six-form is a confusing expression.
However, in order to emphasize the Killing vectors of the isometries,
we adopted this notation.}
\begin{align}
B^{(6)} \ \xrightarrow[k_1 k_2]{} \ 
i_{k_1} i_{k_2} B^{(8,2)}
\, .
\end{align}
Along with the backgrounds fields, the world-volume gauge field
is transformed as 
\begin{align}
\mathcal{A}_a \ \xrightarrow[k_1 k_2]{} \ \wt{A}_a
\, .
\end{align}

Collecting all the formulae together, 
we explicitly write down the world-volume action of
the type IIB exotic $5^2_2$-brane:
\begin{align}
S^{\text{IIB}}_{5^2_2} 
\ =& \ 
- T_{5^2_2} \int \! \d^6 \xi \, \e^{-2\phi} (\det h_{IJ})
\sqrt{
1 + \frac{\e^{2\phi} (i_{k_1} i_{k_2} (C^{(2)} + C^{(0)} B))^2}{\det h_{IJ}}
} 
\nn \\
\ & \ \times
\sqrt{
- \det 
\Big(
\Pi_{\mu \nu} (k_2) \, \partial_a X^{\mu} \partial_b X^{\nu} 
+ \frac{K_a^{(1)} K_b^{(1)}}{(k_2)^2} 
- \frac{(k_2)^2 (K_a^{(2)} K_b^{(2)} - K^{(3)}_a K^{(3)}_b)}{\det h_{IJ}} 
+ \lambda \mathscr{F}_{ab}
\Big)
}
\nn \\
& \ 
- \mu_5 \int_{\mathcal{M}_6} \! \left[
\, P [i_{k_1} i_{k_2} B^{(8,2)}]
- \frac{1}{2} P [ \wt{B} \wedge \wt{C}^{(2)} \wedge \wt{C}^{(2)} ]
- \lambda P [ \wt{C}^{(4)} + \wt{C}^{(2)} \wedge \wt{B} ] 
\wedge \wt{F}
\right.
\nn \\
\ & \LS \
\left. 
\frac{}{}
- \frac{\lambda^2}{2!} 
P [\wt{B}] \wedge \wt{F} \wedge \wt{F}
+ \frac{\lambda^3}{3!} \frac{i_{k_1} i_{k_2} (C^{(2)} + C^{(0)}
 B)}{(i_{k_1} i_{k_2} (C^{(2)} + C^{(0)} B))^2 + \e^{-2\phi} (\det h_{IJ})}
\wt{F} \wedge \wt{F} \wedge \wt{F}
\right]
\, ,
\label{eq:IIB522}
\end{align}
where $T_{5^2_2} = T_{\text{NS5}}$ 
and we have defined 
\begin{align}
\wt{F}_{ab}
\ &= \
\partial_a \wt{A}_b - \partial_b \wt{A}_a 
- \lambda^{-1} P[ \wt{C}^{(2)} ]_{ab}
\, .
\end{align}
The action contains the following quantity:
\begin{align}
\mathscr{F}_{ab}
\ &= \ 
\frac{\e^{\phi}}{\sqrt{\det h_{IJ} + \e^{2\phi}
 (i_{k_1} i_{k_2} (C^{(2)} + C^{(0)} B))^2}} 
\, \wt{F}_{ab}
\, .
\end{align}
The gauge transformation of $i_{k_1} i_{k_2} B^{(8,2)}$ is complicated
but is determined straightforwardly such that the Wess-Zumino term of
the action \eqref{eq:IIB522} is invariant under the following transformations:
\begin{eqnarray}
\begin{aligned}
\delta B \ &= \ 
\d \Lambda^{(1)}
\, , \\
\delta C^{(0)} \ &= \ 0
\, , \ls
\delta C^{(2)} \ = \ 
\d \lambda^{(1)}
\, , \ls
\delta C^{(4)} \ = \ 
\d \lambda^{(3)} - B \wedge \d \lambda^{(1)}
\, , \\
\delta (\d \wt{A}^{(1)})
\ &= \ 
\lambda P [\delta \wt{C}^{(2)}]
\, , \ls
\delta \varphi \ = \ 
- \lambda^{-1} \, i_{k_1} \Lambda^{(1)}
\, , \ls
\delta \varphi' \ = \ 
- \lambda^{-1} \, i_{k_2} \Lambda^{(1)}
\, .
\end{aligned}
\label{eq:IIB522gauge}
\end{eqnarray}
Note that the effective action \eqref{eq:IIB522} is written in the space-time covariant
fashion which is compared with the action obtained in
\cite{Chatzistavrakidis:2013jqa}. 
The action \eqref{eq:IIB522} shows specific properties of the exotic $5^2_2$-brane.
The dilation coupling $\e^{-2\phi}$ implies that the brane is a solitonic
object as its notation stands for.
We also see that the overall factor $\det h_{IJ}$, which contains 
$(k_1)^2 (k_2)^2$, reflects the fact that the tension of the $5^2_2$-brane is
proportional to the radii of the two compactified (isometry) directions.
This is the generalization of the structure seen in the KK-monopole
world-volume \cite{Bergshoeff:1997gy, Eyras:1998hn}
where the tension is proportional to the Killing vector $k^2$ associated with the Taub-NUT isometry.
In order to see that the effective action \eqref{eq:IIB522} contains
zero-modes that correspond to two transverse directions,
we focus on the determinant part of the action. 
When the backgrounds other than the metric are zero and
$\wt{A}_a = \varphi = \varphi' = 0$, the determinant part of the action is 
\begin{align}
& \Pi_{\mu \nu} (k_2)
\partial_a X^{\mu} \partial_b X^{\nu} 
\nn \\
\ & \ \ \ \ 
- \frac{k_2^2}{\det h_{IJ}}
\Big[
(i_{k_1} g)_{\mu} \partial_a X^{\mu} 
- \frac{k_1 \cdot k_2}{(k_2)^2} (i_{k_2} g)_{\mu} \partial_a X^{\mu}
\Big] 
\Big[
(i_{k_1} g)_{\nu} \partial_b X^{\nu} 
- \frac{k_1 \cdot k_2}{(k_2)^2} (i_{k_2} g)_{\nu} \partial_b X^{\nu}
\Big]
\nn \\
\ &= \ 
\Pi_{\mu \nu} \partial_a X^{\mu} \partial_b X^{\nu}
\, ,
\end{align}
where $\Pi_{\mu \nu} = g_{\mu \nu} - h^{IJ} g_{\mu \rho} g_{\nu \sigma}
k^{\rho}_I k^{\sigma}_J$, $h_{IJ} = g_{\mu \nu} k^{\mu}_I k^{\nu}_J$ and
$h^{IK} h_{KJ} = \delta^I {}_J$. 
This is nothing but the structure of the gauged sigma model with two
isometries \cite{Bergshoeff:1997gy}.
The fluctuation modes associated with the isometry directions are
projected out.
We note that when the NS-NS $B$-field is turned on, 
the action is not written as a gauged sigma model.
The exotic seven-brane with two gauged isometries is constructed in
\cite{Eyras:1999at} where the NS-NS $B$-field is turned off.

In the static gauge, the field content of the effective action
\eqref{eq:IIB522} is therefore, a $U(1)$ gauge field $\wt{A}_a$, two
transverse scalar zero-modes $X^1$, $X^2$, two scalar fields $\varphi$,
$\varphi'$ that come from the T-duality transformation. 
They are organized into the six-dimensional $\mathcal{N} = (1,1)$ vector
multiplet as we expected.

Next, we proceed to the exotic $5^2_3$-brane.
The effective action of the exotic $5^2_3$-brane is
obtained by applying the S-duality transformation 
to the $5^2_2$-brane effective action \eqref{eq:IIB522}
\cite{Chatzistavrakidis:2013jqa}
(see also Figure \ref{fig1}).
The world-volume fields are transformed as 
\begin{align}
\wt{A}_a \ \xrightarrow[\text{S}]{} \ \wt{A}^{\prime}_a
\, , \ls 
\varphi \ \xrightarrow[\text{S}]{} \ \wt{\varphi}
\, , \ls 
\varphi' \ \xrightarrow[\text{S}]{} \ \wt{\varphi}'
\, .
\end{align}
The background fields are transformed by the S-duality rule
\eqref{eq:S-duality}. 
Then the action is 
\begin{align}
S^{\mathrm{IIB}}_{5^2_3} 
\ =& \ 
- T_{5^2_3} \int \! \d^6 \xi \, 
|\tau| \, \e^{-2 \phi} (\det l_{IJ})
\sqrt{
1 + \frac{\e^{2\phi}}{\det l_{IJ}} 
\Big\{
i_{k_1} i_{i_2} (B + |\tau|^{-2} C^{(0)} C^{(2)})
\Big\}
}
\nn \\
& \ \times 
\sqrt{
- \det 
\Big(
\Pi_{\mu \nu} (k_2) \, \partial_a X^{\mu} \partial_b X^{\nu} 
+ \frac{L_a^{(1)} L_b^{(1)}}{|\tau|^2 (k_2)^2} 
- \frac{(k_2)^2 (L_a^{(2)} L_b^{(2)} - |\tau|^{-2} L^{(3)}_a L^{(3)}_b)}{\det l_{IJ}} 
+ \lambda \mathscr{F}'_{ab} 
\Big)
}
\notag \\
& \ - \mu_5 \int_{\mathcal{M}_6} \! 
\left[
P[i_{k_1} i_{k_2} C^{(8,2)}] - \frac{1}{2} P[\wt{B}^{\prime} \wedge
 \wt{C}^{(2) \prime} \wedge \wt{C}^{(2) \prime}]
\right. 
\nn \\
& \LS \ls
- \lambda P [\wt{C}^{(4) \prime} \wedge \wt{C}^{(2) \prime} \wedge
 \wt{B}^{\prime}] \wedge \wt{F}^{\prime}
- \frac{\lambda^2}{2!} P[\wt{B}^{\prime}] 
\w \wt{F}^{\prime} \wedge \wt{F}^{\prime} 
\nn \\
& \LS \ls
+ \frac{\lambda^3}{3!}
\frac{|\tau|^2 \, \e^{2\phi} \, i_{k_1} i_{k_2} (B + |\tau|^{-2} C^{(0)}
 C^{(2)})}{\det l_{IJ} + |\tau|^2 \, \e^{2\phi} (i_{k_1} i_{k_2} (B + |\tau|^{-2} C^{(0)} C^{(2)} ))^2}
 \wedge \wt{F}^{\prime} \wedge \wt{F}^{\prime} \wedge \wt{F}^{\prime}
\biggr]
\, ,
\label{eq:IIB523}
\end{align}
where $T_{5^2_3} = g^{-1}_{\text{st}} T_{5^2_2}$ and 
we have defined 
\begin{eqnarray}
\begin{aligned}
l_{IJ} \ &= \ 
k^{\mu}_I k^{\nu}_J (g_{\mu \nu} - |\tau|^{-1} C^{(2)}_{\mu \nu})
\, , \\
L^{(1)}_{\mu} 
\ &= \ 
(i_{k_2} C^{(2)} + \lambda \d \wt{\varphi}')_{\mu}
\, , \\
L^{(2)}_{\mu} 
\ &= \ 
\Big(
i_{k_1} g - \frac{k_1 \cdot k_2}{(k_2)^2} (i_{k_2} g) 
+ \frac{1}{|\tau|^2 (k_2)^2} (i_{k_1} i_{k_2} C^{(2)}) 
(i_{k_2} C^{(2)} + \lambda \d \wt{\varphi}')
\Big)_{\mu}
\, , \\
L^{(3)}_{\mu} \ &= \ 
\Big(
(i_{k_1} C^{(2)} + \lambda \d \wt{\varphi}) 
- \frac{k_1 \cdot k_2}{(k_2)^2} (i_{k_2} C^{(2)} + \lambda \d 
\wt{\varphi}'
) 
+ \frac{1}{(k_2)^2} (i_{k_1} i_{k_2} C^{(2)}) i_{k_2} g
\Big)_{\mu}
\, , \\
\wt{F}'_{ab} 
\ &= \ 
\partial_a \wt{A}'_b - \partial_b
 \wt{A}'_a - \lambda^{-1} P[\wt{C}^{(2) \prime}]_{ab}
\, , \\
\mathscr{F}'_{ab} 
\ &= \ 
\frac{\e^{\phi}}
{\sqrt{\det l_{IJ} + |\tau|^2 \, \e^{2\phi} i_{k_1} i_{k_2} (B + |\tau|^{-2} C^{(0)} C^{(2)})}}
\, \wt{F}'_{ab}
\, .
\end{aligned}
\end{eqnarray}
The explicit forms of the background fields $\wt{B}^{\prime}$,
$\wt{C}^{(2) \prime}$, $\wt{C}^{(4) \prime}$ are found in
\eqref{eq:523backgrounds}.
The power of the dilaton field implies that the tension of this five-brane is proportional to $g_{\text{st}}^{-3}$.
Notice that $C^{(8,2)}$ is the R-R mixed-symmetry tensor.
As we addressed in Section \ref{sect-review522}, 
this is S-dual of the NS-NS mixed-symmetry tensor $B^{(8,2)}$ in ten
dimensions \cite{Chatzistavrakidis:2013jqa}. 
Although the $5^2_3$-brane does not exist in type IIA string theory, 
one encounters this exotic brane in type I theory.
We will come back to the $5^2_3$-brane in Section \ref{sect-WVI}.

\subsection{$5^2_2$-brane in type IIA theory}
In this subsection, we consider the exotic $5^2_2$-brane in type IIA
string theory. The supermultiplet of the type IIA $5^2_2$-brane
effective theory is a six-dimensional $\mathcal{N} = (2,0)$ tensor multiplet which consists of one self-dual tensor field of rank two and five scalar fields.
It is necessary to introduce an auxiliary field for a manifestly Lorentz
invariant action of self-dual fields.
In order to obtain the IIA $5^2_2$-brane, we start from the action of
a single M5-brane in eleven dimensions.
By the direct dimensional reduction of the M5-brane,
we first obtain the type IIA NS5-brane.
Performing the T-duality transformations twice, we obtain the type IIA $5^2_2$-brane.

The world-volume fields in the M5-brane effective action belong to a six-dimensional $\mathcal{N} = (2,0)$ tensor multiplet. 
The five scalar fields correspond to
the translational moduli of the M5-brane in eleven dimensions. 
The action involves the $U(1)$ self-dual two-form gauge field $A_{ab}$.
The manifestly Lorentz invariant form of the action is given in the Pasti-Sorokin-Tonin (PST) formalism \cite{Pasti:1997gx}:
\begin{align}
S_{\text{M5}} 
\ =& \ 
- T_{\text{M5}} \int\! \d^6 \xi \,
\left[
\sqrt{
- \det (P[\wh{g}]_{ab} + \i \wh{H}^{*}_{ab})
}
+ \frac{\sqrt{- \wh{g}}}{4 (\wh{\del a})^2}
\wh{H}^{*abc} \wh{H}_{bcd} \, (\partial_a a \, \partial^d a)
\right]
\nn \\
\ & \ 
+ T_{\text{M5}} \int_{\mathcal{M}_6} \! 
\left(
P[\wh{C}^{(6)}] - \frac{1}{2} F^{(3)} \wedge P[\wh{C}^{(3)}]
\right)
\, ,
\label{eq:M5}
\end{align}
where $T_{\text{M5}} = \frac{1}{(2\pi)^5} M^6_{11}$ is the M5-brane tension in terms of the the Planck mass $M_{11}$ in eleven dimensions.
$F^{(3)} = \d A^{(2)}$ is the field strength of the self-dual two-form gauge field.
The real auxiliary field $a(\xi)$ is non-dynamical. 
The explicit expressions of the variables in the action are given by 
\begin{eqnarray}
\begin{aligned}
F_{abc}^{(3)} 
\ &= \ 
\partial_a A_{bc} - \partial_b A_{ac} + \partial_c A_{ab}
\, , \ls 
\wh{g} \ = \ \det P[\wh{g}]
\, , \ls 
(\wh{\del a})^2 \ = \ 
P[\wh{g}]^{ab} \, \partial_a a \, \partial_b a
\, , \\
\wh{H}_{abc} \ &= \ 
F_{abc}^{(3)} - P[\wh{C}^{(3)}]_{abc}
\, , \quad
\wh{H}^{*abc} \ = \ 
\frac{1}{3! \sqrt{-\wh{g}}} \, \varepsilon^{abcdef} \, \wh{H}_{def}
\, , \quad
\wh{H}^{*}_{ab} \ = \ 
\frac{1}{\sqrt{(\wh{\del a})^2}} \, \wh{H}^{*}_{abc} \, \partial^c a
\, .
\end{aligned}
\end{eqnarray}
All the world-volume indices are contracted with $P[g]_{ab}$, {\it
i.e.}, the pull-back $P[\wh{g}]^{ab}$.
Here $P[\wh{g}]^{ab}$ is the inverse of $P[\wh{g}]_{ab}$.
The fields
$\wh{g}_{MN}$, $\wh{C}^{(3)}$, $\wh{C}^{(6)}$ are the space-time metric, 
the three-form potential and its magnetic dual in $D=11$ supergravity. 
The action is invariant under the following four kinds of gauge transformations:
\\
\\
\textbullet \ The ordinary world-volume gauge transformation 
by the gauge parameter one-form $\chi_a$:
\begin{align}
\delta A_{ab} 
\ &= \ 
\partial_{[a} \chi_{b]}
\, .
\label{eq:M5gauge1}
\end{align}
\textbullet \ The field dependent world-volume gauge transformation
by the gauge parameter one-form $\chi'_a$:
\begin{align}
\delta A_{ab} 
\ &= \ 
\partial_{[a}^{\vphantom{\prime}} a \ \chi'_{b]}
\, , \ls 
\delta a \ = \ 0
\, .
\label{eq:M5gauge2}
\end{align}
\textbullet \ 
The field dependent world-volume gauge transformation
by the gauge parameter $\wh{\chi}$:
\begin{eqnarray}
\begin{aligned}
\delta a \ &= \ 
\wh{\chi}
\, , \ls
\delta A_{ab} \ = \ 
\frac{\wh{\chi}}{2 (\wh{\del a})^2} 
(H_{abc} \, \partial^c a - \mathcal{V}_{ab})
\, , \\
\mathcal{V}^{ab} 
\ &= \ 
- 2 
\sqrt{ - \frac{(\wh{\del a})^2}{\wh{g}}} \frac{\delta}{\delta \wh{H}^{*}_{ab}} 
\sqrt{- \det (P[\wh{g}] + \i \wh{H}^{*})}
\, .
\end{aligned}
\label{eq:M5gauge3}
\end{eqnarray}
\textbullet \ 
The space-time gauge transformations by the gauge parameter five- and two-forms $\wh{\Lambda}^{(5)}$, $\wh{\Lambda}^{(2)}$:
\begin{eqnarray}
\begin{gathered}
\delta \wh{C}^{(3)} (X) 
\ = \ 
\d \wh{\Lambda}^{(2)} (X)
\, , \ls
\delta \wh{C}^{(6)} (X) 
\ = \ 
\d \wh{\Lambda}^{(5)} + 
\frac{1}{2} \d \wh{\Lambda}^{(2)} (X) \wedge \wh{C}^{(3)} (X)
\, , \\
\delta A^{(2)} (\xi) 
\ = \ 
\wh{\Lambda}^{(2)} (X(\xi))
\, . 
\end{gathered}
\end{eqnarray}

Now we perform the direct dimensional reduction of the M5-brane action
\eqref{eq:M5}. 
The KK ansatz is 
\begin{align}
\wh{g}_{MN} 
\ &= \ 
\left(
\begin{array}{cc}
\e^{-\frac{2}{3} \phi} (g_{\mu \nu} + \e^{2\phi} \, C^{(1)}_{\mu} C^{(1)}_{\nu} ) 
& \e^{\frac{4}{3} \phi} \, C^{(1)}_{\mu} 
\\
\e^{\frac{4}{3} \phi} \, C^{(1)}_{\nu} & \e^{\frac{4}{3} \phi}
\end{array}
\right)
\, , 
\label{eq:KKansatz}
\end{align}
where $M,N=0, \ldots, 10$ is the space-time indices in eleven dimensions.
The potential forms are decomposed as 
\begin{eqnarray}
\begin{aligned}
\wh{C}^{(3)} 
\ &= \ 
C^{(3)} - B \wedge \d Y
\, , \\
\wh{C}^{(6)} 
\ &= \ 
B^{(6)} + C^{(5)} \wedge \d Y 
+ \frac{1}{2} C^{(5)} \wedge C^{(1)} 
+ \frac{1}{2} C^{(3)} \wedge B \wedge \d Y
\, .
\end{aligned}
\end{eqnarray}
Here $Y$ is a world-volume scalar field associated with the compact
space-time coordinate $y$.
Note that the fields in the left-hand sides are form fields in $D=11$
while the ones in the right-hand sides are those in $D=10$.
For later convenience,
we also define
\begin{align}
F_{\mu} \ = \ 
\partial_{\mu} Y + C^{(1)}_{\mu}
\, .
\end{align}
Performing the direct dimensional reduction, we obtain the effective
action of the type IIA NS5-brane \cite{Bandos:2000az}:
\begin{align}
S^{\text{IIA}}_{\text{NS5}}
\ =& \
- T_{\text{NS5}} \int \! \d^6 \xi \, \e^{-2\phi} 
\sqrt{
- \det (P[g]_{ab} + \lambda^2 \e^{2 \phi} F_a F_b) 
}
\sqrt{
\det 
\Big(
\delta_a{}^b + \frac{\i \lambda \, \e^{\phi} (P[g]_{ac} + \lambda^2 \e^{2\phi} F_a F_c)}{\mathcal{N} \sqrt{\det (\delta_{e}{}^f + \lambda^2 \e^{2\phi} F_e F^f)}}
\, H^{* bc}
\Big)
}  
\nn \\
\ & \
- \frac{\lambda^2}{4} T_{\text{NS5}} \int \! \d^6 \xi \, 
\frac{\sqrt{-g}}{\mathcal{N}^2} \, H^{* ab} H_{abc} 
\Big( P[g]^{cd} - \frac{\e^{2\phi} \lambda^2 F^c F^d}{1 + \lambda^2 \e^{2\phi} F^2}
 \Big) 
\frac{\partial_d a}{\sqrt{(\partial a)^2}}
\nn \\
\ & \
+ \mu_5 \int_{\mathcal{M}_6} \! 
\left(
P[B^{(6)}] + \lambda P[C^{(5)} \wedge \d Y] + \frac{1}{2} P [C^{(5)} \wedge C^{(1)}] 
+ \frac{\lambda}{2} P[C^{(3)} \wedge B \wedge \d Y]
\right.
\nn \\
\ & \LS \ls \
\left.
- \frac{\lambda}{2} F^{(3)} \wedge P[C^{(3)}] 
+ \frac{\lambda}{2} F^{(3)} \wedge P[B \wedge \d Y]
\right)
\, ,
\label{eq:IIANS5}
\end{align}
where we have defined 
\begin{eqnarray}
\begin{aligned}
F_a \ &= \ 
P[F]_a \ = \ \partial_a Y + \lambda^{-1} P[C^{(1)}]_a
\, , \ls
\mathcal{N} 
\ = \
\sqrt{
1 - \frac{\lambda^2 \e^{2\phi} (F \partial a)^2}{(\partial a)^2 (1 + \lambda^2 \e^{2\phi} F^2)}
}
\, , \\
g \ &= \ \det P[g]
\, , \ls
(\del a)^2 
\ = \ 
P [g]^{ab} \, \del_a a \, \del_b a
\, , \\
H_{abc} \ &= \ 
F^{(3)}_{abc} - \lambda^{-1} P[C^{(3)}]_{abc} - (P[B] \wedge \d Y)_{abc}
\, , \\
H^{* abc} \ &= \ \frac{1}{3! \sqrt{-g}} \, \varepsilon^{abcdef} H_{def}
\, , \ls
H^{*ab} \ = \ \frac{1}{\sqrt{(\partial a)^2}} H^{*abc} \, \partial_c a
\, .
\end{aligned}
\end{eqnarray}
Here we have rescaled 
$F^{(3)}_{abc} \to \lambda F^{(3)}_{abc}$, $Y \to \lambda Y$ 
in order to make the dimension of the fields be $[A_{ab}] = [Y] = +1$. 
All the world-volume indices are contracted with the pull-back of the ten-dimensional metric $g_{\mu \nu}$.
The Wess-Zumino term is invariant under the following space-time gauge transformations:
\begin{eqnarray}
\begin{aligned}
\delta B \ &= \ \d \Lambda^{(1)}
\, , \\
\delta B^{(6)} \ &= \ 
\d \Lambda^{(5)} + \frac{1}{2} C^{(1)} \wedge \d \lambda^{(4)} 
- \frac{1}{2} \Big( C^{(3)} + C^{(1)} \wedge B \Big) \wedge \d \lambda^{(2)} 
\\
\ & \ \ \ \ 
+ \frac{1}{2} \Big(
C^{(5)} + C^{(3)} \wedge B 
+ \frac{1}{2} C^{(1)} \wedge B \wedge B 
\Big) \wedge \d \lambda^{(0)}
\, , \\
\delta C^{(1)} \ &= \ 
\d \lambda^{(0)}
\, , \\
\delta C^{(3)} \ &= \ 
\d \lambda^{(2)} - B \wedge d \lambda^{(0)}
\, , \\
\delta C^{(5)} \ &= \ 
\d \lambda^{(4)} 
- B \wedge \d \lambda^{(2)} 
+ \frac{1}{2!} B \wedge B \wedge \d \lambda^{(0)}
\, , \\
\delta Y \ &= \ 
- \lambda^{(0)}
\, , \\
\delta (\d A^{(2)}) \ &= \ 
\lambda^{-1} P [\d \lambda^{(2)} - B \wedge \d \lambda^{(0)}] 
+ P [\d \Lambda^{(1)} \wedge \d Y - B \wedge \d \lambda^{(0)}]
\, ,
\end{aligned}
\end{eqnarray}
where $\Lambda^{(n)}$ and $\lambda^{(n)}$ are gauge parameter $n$-forms.

Now we perform the T-duality transformations of the action
\eqref{eq:IIANS5}. 
The pull-back of the metric is transformed as 
\begin{eqnarray}
\begin{aligned}
P[g]_{ab} 
\ &\xrightarrow[k_1 k_2]{} \
\Pi_{\mu \nu} (k_2)
\, \partial_a X^{\mu} \partial_b X^{\nu}
+ \frac{K_a^{(1)} K_b^{(1)}}{(k_2)^2} 
- \frac{(k_2)^2 (K_a^{(2)} K_b^{(2)} - K_a^{(3)} K_b^{(3)})}{\det h_{IJ}} 
\\
\ &\ \ \equiv \ \
\wt{g}_{ab}
\, .
\end{aligned}
\end{eqnarray}
We define the inverse matrix $\wt{g}^{ab}$ as follows:
\begin{align}
\wt{g}_{ab} \, \wt{g}^{bc} 
\ &= \ 
\wt{g}^{cb} \, \wt{g}_{ba} 
\ = \ 
\delta_a {}^c
\, .
\end{align}
The T-duality transformations of the R-R fields are 
\begin{gather}
P[C^{(1)}] \ \xrightarrow[k_1 k_2]{} \ P[\wt{C}^{(1)}]
\, , \ls
P[C^{(3)}] \ \xrightarrow[k_1 k_2]{} \ P[\wt{C}^{(3)}]
\, , \ls
P[C^{(5)}] \ \xrightarrow[k_1 k_2]{} \ P[\wt{C}^{(5)}]
\, ,
\end{gather}
where the explicit expressions of the right-hand sides are found in Appendix \ref{app-covBuscher}.
We also have 
\begin{eqnarray}
\begin{aligned}
P[g]_{ab} + \e^{2\phi} F_a F_b 
\ &\xrightarrow[k_1 k_2]{} \ 
\wt{g}_{ab} 
+ \frac{\e^{2\phi}}{\det h_{IJ}}
(\partial_a Y + P[\wt{C}^{(1)}]_a)
(\partial_b Y + P[\wt{C}^{(1)}]_b)
\, , \\
F^a \partial_a a 
\ &\xrightarrow[k_1 k_2]{} \ 
\wt{g}^{ab} (\partial_a Y + P[\wt{C}^{(1)}]_a) \, \partial_b a
\, , \\
F^2 \ &\xrightarrow[k_1 k_2]{} \ 
\wt{g}^{ab} (\partial_a Y + P[\wt{C}^{(1)}]_a)
(\partial_b Y + P[\wt{C}^{(1)}]_b)
\, , \\
\mathcal{N}^2 \ &\xrightarrow[k_1 k_2]{} \ 
1 - \frac{\e^{2\phi} \, \wt{g}^{ef} (\partial_e Y + P[\wt{C}^{(1)}]_e)
\, \partial_f a}{(\wt{\del a})^2
\left(
\det h_{IJ} + \e^{2\phi} \wt{g}^{cd} 
(\partial_c Y + P[\wt{C}^{(1)}]_c)
(\partial_d Y + P[\wt{C}^{(1)}]_d)
\right)
}
\\
\ &\ \ \equiv \ \
\wt{\mathcal{N}}^2
\, ,
\end{aligned}
\end{eqnarray}
where $(\wt{\del a})^2 = \wt{g}^{ab} \del_a a \, \del_b a$, and
\begin{align}
H_{abc} \ &\xrightarrow[k_1 k_2]{} \ 
\wt{F}_{abc} - P[\wt{C}^{(3)}]_{abc} - (P[\wt{B}] \wedge 
(\d Y + P[\wt{C}^{(1)}]))_{abc} 
\ \equiv \ \wt{H}_{abc}
\, .
\end{align}
Here the world-volume gauge field $A_{ab}$ is transformed as 
$A_{ab} \xrightarrow[k_1 k_2]{} \wt{A}_{ab}$ and $\wt{F}^{(3)} = \d \wt{A}^{(2)}$. 
Putting everything together, 
we obtain the effective action of the type IIA $5^2_2$-brane:
\begin{align}
& S^{\mathrm{IIA}}_{5^2_2} 
\nn \\
\ &= \ - T_{5^2_2} \int \! \d^6 \xi \, 
\e^{-2\phi} (\det h_{IJ}) 
\nn \\
& \ \ \ \ \times 
\sqrt{
- \det 
\Big(
\Pi_{\mu \nu} (k_2)
\partial_a X^{\mu} \partial_b X^{\nu}
+ \frac{K_a^{(1)} K_b^{(1)}}{(k_2)^2} 
- \frac{(k_2)^2 (K_a^{(2)} K_b^{(2)} - K_a^{(3)} K_b^{(3)})}{\det h_{IJ}}
 + \frac{\lambda^2 \e^{2\phi}}{\det h_{IJ}}
\wt{F}^{(1)}_a \wt{F}^{(1)}_b
\Big)
} 
\nn \\
& \ \ \ \ \times 
\sqrt{
\det 
\Big(
\delta_a{}^b 
+ \frac{\i \, \e^{\phi}}{3! \, \wt{\mathcal{N}} \sqrt{\det h_{IJ} (\wt{\del a})^2}} Z_a{}^b
\Big)
}
\nn \\
& \ \ \ \ 
- \frac{\lambda^2}{4} T_{5^2_2} 
\int \! \d^6 \xi \, 
\frac{\varepsilon^{abgd'e'f'} \, \wt{H}_{d'e'f'} \, \wt{H}_{abc} \, (\partial_g a \, \partial_d a)}{3! \, \wt{\mathcal{N}}^2 (\wt{\del a})^2} 
\Bigg[
\wt{g}^{cd} 
- \frac{\lambda^2 \e^{2\phi} \, \wt{g}^{ce} \, \wt{g}^{df} \,
\wt{F}^{(1)}_e \wt{F}^{(1)}_f 
}{\det
 h_{IJ} + \lambda^2 \e^{2\phi} \, \wt{g}^{a'b'} \, \wt{F}^{(1)}_{a'} \wt{F}^{(1)}_{b'}
}
\Bigg]
\nn \\
& \ \ \ \ 
+ T_{5^2_2} \int_{\mathcal{M}_6} \! 
\left(
P[i_{k_1} i_{k_2} B^{(8,2)}] 
+ \frac{1}{2} P [\wt{C}^{(5)} \wedge \wt{C}^{(1)} ]
+ \lambda P[\wt{C}^{(5)}] \wedge \d Y
+ \frac{\lambda}{2} P[\wt{C}^{(3)} \wedge \wt{B} \wedge \d Y]
\right. 
\nn \\
\ & \LS \LS \ \
\left.
- \frac{\lambda}{2} F^{(3)} \wedge P [\wt{C}^{(3)}]
+ \frac{\lambda}{2} F^{(3)} \wedge P [\wt{B} \wedge \d Y]
\right)
\, .
\label{eq:IIA522}
\end{align}
Here we introduced the following expressions:
\begin{eqnarray}
\begin{aligned}
\wt{F}^{(1)}_a 
\ &= \ 
\partial_a Y + \lambda^{-1} P[\wt{C}^{(1)}]_a
\, , \\
Z_a{}^b 
\ &= \ 
\frac{\varepsilon^{gbcdef} \big( \wt{g}_{ac} + \lambda^2 \e^{2\phi} (\det h_{IJ})^{-1} \,
\wt{F}^{(1)}_a \wt{F}^{(1)}_b \big) \, \wt{H}_{def} \, \partial_g a}
{\sqrt{- \det \big( \wt{g}_{a'b'} + \lambda^2 \, \e^{2\phi} (\det h_{IJ})^{-1} 
\wt{F}^{(1)}_{a'} \wt{F}^{(1)}_{b'}
\big)
}}
\, .
\end{aligned}
\end{eqnarray}
The gauge transformations of the scalar fields $\varphi, \varphi'$
are those found in \eqref{eq:IIB522gauge}. 
The gauge transformation of the scalar field $Y$ is defined such that
the modified field strength $\wt{F}^{(1)}$ is invariant, namely, 
\begin{align}
\delta (\d Y) \ = \ - \lambda^{-1} P[\delta \wt{C}^{(1)}]
\, .
\end{align}
The world-volume gauge transformations of $\wt{A}_{ab}$ and
the auxiliary field $a$ are obtained from
those in the M5-brane \eqref{eq:M5gauge1}-\eqref{eq:M5gauge3} where the
background fields are given in the KK ansatz \eqref{eq:KKansatz} with
the T-dualized forms.
Again the space-time gauge transformations of the non-geometric flux
$i_{k_1} i_{k_2} B^{(8,2)}$ is determined by requiring that the
Wess-Zumino term of the action \eqref{eq:IIA522} is invariant under the gauge transformations of
$\wt{C}^{(5)}$, $\wt{C}^{(3)}$, $\wt{C}^{(1)}$ and $\wt{B}$. 
The calculation is straightforward but tedious and we never pursue it here.
Now, we summarize the field content of the effective action
\eqref{eq:IIA522}. 
In the static gauge, the world-volume fields are 
two translational zero-modes $X^1$,
$X^2$, a scalar field $Y$ originated from the dimensional reduction from
$D=11$ to $D=10$, two scalar fields $\varphi$, $\varphi'$ associated
with the dual coordinates, a self-dual two-form gauge field $\wt{A}_{ab}$ and
a non-dynamical auxiliary field $a$.
They are organized into the six-dimensional $\mathcal{N} = (2,0)$ tensor
multiplet as expected.

A comment is in order about world-volume effective actions of 
exotic five-branes.
As displayed in Figure \ref{fig1}, there are other exotic five-branes of
co-dimension less than two in type II string theories.
For example, performing the T-duality transformation of the
$5^2_3$-brane, we obtain another exotic five-brane called
$5^{(1,2)}_3$-brane in type IIA theory.
This has the tension proportional to $g_{\text{st}}^{-3}$
and is a heavy object in perturbative string theory.
The world-volume effective action of the $5^{(1,2)}_3$-brane should contain three
Killing vectors $k^{\mu}_1$, $k^{\mu}_2$, $k^{\mu}_3$ and 
the supermultiplet of the theory must be the six-dimensional
$\mathcal{N} = (2,0)$ tensor multiplet.
Although it is formally possible to write down the effective action of the
exotic five-branes of co-dimension less than two in the covariant
fashion, we never explore it in this paper.

\section{Exotic five-branes in type I and heterotic string theories}
\label{sect-WVI}

In the previous section, we gave the world-volume effective actions 
of the exotic $5^2_2$- and $5^2_3$-branes in type II string theories by
virtue of the string duality chains.
In this section, we explore exotic five-branes in type I and heterotic string theories. 
Through the string duality chains, 
we can also discuss the effective actions of the exotic five-branes in these theories.

The most notable fact about the world-volume theory of five-branes in
type I and heterotic theories is its non-Abelian nature even for
a single brane.
For example, the type I single D5-brane 
supports the $Sp(1) = SU(2)$ gauge group in its world-volume
\cite{Angelantonj:2002ct}.
In addition to the vector multiplet, there are hypermultiplets transformed as $(\mathbf{2},\mathbf{32})$
of the $SU(2) \times SO(32)$ group.
One consequence of this fact is that exotic five-branes in heterotic
theories should have non-Abelian gauge groups by the duality chains 
(see Figure \ref{fig2}).
It is a common practice that non-Abelian gauge groups in 
the DBI type of the actions are hard to analyze.
We therefore consider a $U(1)$ subgroup of the gauge groups in the effective actions and extract some physical properties of the exotic five-branes in type I and heterotic theories.

\subsection{$5^2_3$-brane in Type I theory}

Type I string theory is obtained by the orientifold projection of type
IIB string theory with the addition of 
an O9-plane, sixteen D9-branes and their mirror images.
The most familiar five-brane in type I theory is the D5-brane.
The D5-brane preserves eight supercharges.
In type I theory, the D5-brane world-volume fields originate from
the D5-D5 and the D5-D9 open string sectors, 
where the D5-D5 sector provides the $Sp(1) = SU(2)$ Chan-Paton factor.
An $\mathcal{N} = (1,0)$ vector and an $\mathcal{N} = (1,0)$
hypermultiplet come from the D5-D5 open string sector.
The vector multiplet belongs to the adjoint representation of $SU(2)$
while the hypermultiplet is $SU(2)$ singlet.
>From the D5-D9 open string sector, there appears to be $SO(2) \times SO(32)$ bi-fundamental
hypermultiplets.

The exotic five-brane in type I theory is the $5^2_3$-brane.
Through the connection to the D5-brane by dualities, 
there are at least one $SU(2)$ adjoint
vector field and four $SU(2)$ singlet real scalar fields
in the world-volume theory of the type I $5^2_3$-brane.
Its $U(1)$ part of the effective action can be obtained from that of the type IIB 
$5^2_3$-brane \eqref{eq:IIB523} by the orientifold projection.
The dilaton $\phi$, the metric $g_{\mu \nu}$, the R-R potentials
$C^{(2)}$, $C^{(6)}$, and the non-geometric fluxes that are related to
these fields by the duality chains, survive the projection.
The half of sixteen supersymmetries is preserved in the world-volume of
the $5^2_3$-brane.
As a result, the $U(1)$ part of the type I $5^2_3$-brane effective
action is
\footnote{
Here we do not consider the world-volume coupling to the background
$SO(32)$ gauge field.
}
\begin{align}
S^{\text{I}}_{5^2_3}
\ &= \ 
- T_{5^2_3} \int \d^6 \xi \, \e^{- 3 \phi} (\det l_{IJ})
\nn \\
\ & \ \ \ \ 
\times \sqrt{
- \det \Big(
\Pi_{\mu \nu} (k_2) \, \del_a X^{\mu} \del_b X^{\nu}
+ \frac{\e^{2 \phi} L_a^{(1)} L_b^{(1)}}{(k_2)^2}
- \frac{(k_2)^2 ( L_a^{(2)} L_b^{(2)} - \e^{2 \phi} L_a^{(3)} L_b^{(3)})}{\det l'_{IJ}}
+ \frac{\lambda \, \e^{\phi}}{\sqrt{\det l'_{IJ}}} \, F'_{ab}
\Big)
}
\nn \\
\ & \ \ \ \ 
+ \mu_5 \int_{\mathcal{M}_6}
\left(
P [i_{k_1} i_{k_2} C^{(8,2)}]
- \frac{\lambda^2}{2} P [\wt{B}^{(2) \prime}] \w F' \w F' 
\right)
\, ,
\end{align}
where we have defined
\begin{eqnarray}
\begin{aligned}
l'_{IJ} \ &= \ 
k_I^{\mu} k_J^{\nu} \big( g_{\mu \nu} - \e^{\phi} C^{(2)}_{\mu \nu} \big)
\, , \\
L_{\mu}^{(1)}
\ &= \ 
(i_{k_2} C^{(2)} + \lambda \d \wt{\varphi}')_{\mu}
\, , \\
L_{\mu}^{(2)}
\ &= \ 
\Big( i_{k_1} g 
- \frac{k_1 \cdot k_2}{(k_2)^2} (i_{k_2} g)
\Big)_{\mu}
+ \frac{\e^{\phi}}{(k_2)^2} (i_{k_1} i_{k_2} C^{(2)})
(i_{k_2} C^{(2)} + \lambda \d \wt{\varphi}')_{\mu}
\, , \\
L_{\mu}^{(3)}
\ &= \ 
\Big(
(i_{k_1} C^{(2)} + \lambda \d \varphi)
- \frac{k_1 \cdot k_2}{(k_2)^2} (i_{k_2} C^{(2)} + \lambda \d \varphi')
+ \frac{1}{(k_2)^2} (i_{k_1} i_{k_2} C^{(2)}) (i_{k_2} g)
\Big)_{\mu}
\, , \\
F'_{ab} \ &= \ \partial_a A'_b - \partial_b A'_a.
\end{aligned}
\end{eqnarray}
The field content of this effective theory is 
a $U(1)$ gauge field $A'_a$, 
two translational zero-modes $X^1$, $X^2$, 
two scalar fields $\wt{\varphi}$, $\wt{\varphi}'$ associated with the dual coordinates. 
Altogether, they consist of an $\mathcal{N} = (1,0)$ Abelian vector
multiplet and a neutral hypermultiplet in six dimensions.
Note that we do not present hypermultiplet parts of the action which are
traced back to the D5-D9 open string sector in the D5-brane.
Although it is intractable to include these parts in the DBI action,
they are necessary for the anomaly free theory of the type I $5^2_3$-brane.

\subsection{$5^2_2$-branes in heterotic theories}

We begin with the half-BPS NS5-brane (heterotic five-brane) in $E_8 \times E_8$ 
or $SO(32)$ heterotic string theory.
In heterotic supergravities, they are gauge instantons embedded in the ten-dimensional space-time geometry \cite{Callan:1991dj, Callan:1991ky}.
The world-volume theory of the heterotic five-brane contains an
$\mathcal{N} = (1,0)$ vector multiplet ($SO(32)$)
or an $\mathcal{N} = (1,0)$ tensor multiplet ($E_8 \times E_8$)
together with an appropriate number of hypermultiplets.
The hypermultiplets are necessary for chiral anomaly free theories.
It is known that the heterotic gauge 
symmetry $E_8 \times E_8$ or $SO(32)$ is broken to $E_7 \times E_8$ 
or $SO(28,2)$ respectively on a heterotic five-brane\footnote{The broken gauge symmetries are obtained by the standard embedding of $SU(2)$ to the original heterotic symmetries. 
Since the $SU(2)$ symmetry comes from the spin connection
\cite{Callan:1991dj, Callan:1991ky}, the symmetry cannot remain as a gauge symmetry. 
Relations between the $SU(2)$ symmetry originated from
spin connections in heterotic string theories and the $SU(2)$ gauge
symmetry of the D5-brane in type I theory are still unclear.},
which is grounded on the anomaly cancellation in six dimensions 
\cite{Izquierdo:1993st, Blum:1993yd, Witten:1995gx, Ganor:1996mu, Mourad:1997uc, Kimura:2009hx, Kimura:2009tb, Imazato:2010qz, Mizoguchi:2012mv}. 
Heterotic string theories contain also exotic $5^2_2$-branes.
Non-geometric solutions in heterotic theories have been discussed in
\cite{Andriot:2011iw, Flournoy:2004vn}.
Through the string dualities,
it is considered that such $E_7 \times E_8$ or $SO(28,2)$ non-Abelian
gauge symmetry must exist on the heterotic $5^2_2$-brane. 
In the following we consider relatively simple part of the effective
actions of the $5^2_2$-brane in heterotic theories.
We appraise that the heterotic $5^2_2$-brane effective actions 
discussed in this section represent an Abelian part of the non-Abelian
gauge symmetries of the fully consistent five-brane world-volume action.
This would substantially lead to consider the neutral solution of heterotic
five-branes \cite{Callan:1991ky} where the $SO(32)$ or $E_8 \times E_8$
gauge field has vanishing configuration.

In $SO(32)$ heterotic string theory, the $5^2_2$-brane effective action
 can be obtained from the type I $5^2_3$-brane effective action by the S-duality transformation:
\begin{align}
C^{(2)} \ \xrightarrow[\text{S}]{} \ B
\, , \ls
\phi \ \xrightarrow[\text{S}]{} \ - \phi
\, , \ls
g_{\mu \nu} \ \xrightarrow[\text{S}]{} \ \e^{-\phi} g_{\mu \nu}
\, , \ls
A^{\text{I}}_{\mu} \ \xrightarrow[\text{S}]{} \ A^{\text{HSO}}_{\mu}
\, ,
\end{align}
where the variables in the left-hand side (right-hand side) are the ones
in type I 
($SO(32)$ heterotic) string theory.
As a result, the $U(1)$ part of the $SO(32)$ heterotic $5^2_2$-brane effective action is
\begin{align}
S^{\text{HSO}}_{5^2_2}
\ &= \ 
- T_{5^2_2} \int \d^6 \xi \, 
\e^{-2 \phi} (\det h_{IJ})
\nn \\
\ & \ \ \ \
\times \sqrt{
- \det \Big(
\Pi_{\mu \nu} (k_2) \, \del_a X^{\mu} \del_b X^{\nu}
+ \frac{K_a^{(1)} K_b^{(1)}}{(k_2)^2}
- \frac{(k_2)^2 (K_a^{(2)} K_b^{(2)} - K_a^{(3)} K_b^{(3)})}{\det h_{IJ}}
+ \frac{\lambda \, \e^{\phi}}{\sqrt{\det h_{IJ}}} \, F_{ab}
\Big)
}
\nn \\
\ & \ \ \ \ 
+ \mu_5 \int_{\mathcal{M}_6} 
\left(
P [i_{k_1} i_{k_2} B^{(8,2)}]
+ \frac{\lambda}{2} F^{(3)} \w P [\wt{B} \w \d Y]
\right)
\, .
\label{eq:SO522}
\end{align}
It is also true that the same effective action is obtained 
by performing two T-duality transformations along two isometry
directions on the $SO(32)$ heterotic five-brane action.
The latter is derived 
via the S-duality transformation of the type I D5-brane effective action. 

We can evaluate the $5^2_2$-brane effective action in 
$E_8 \times E_8$ heterotic theory 
through the repeated T-duality transformations from that of the 
$SO(32)$ heterotic five-brane.
First, we perform a T-duality transformation along the world-volume
direction of the $SO(32)$ heterotic five-brane in order to obtain the
$E_8 \times E_8$ heterotic five-brane effective action.
The transformation rule is the same one between the type IIA and IIB
NS5-branes \cite{Eyras:1998hn}.
Next, we perform two T-duality transformations along transverse directions of the $E_8 \times E_8$ heterotic five-brane. 
Then, the $U(1)$ part of the $E_8 \times E_8$ heterotic $5^2_2$-brane
effective action emerges as
\begin{align}
S^{\text{HE8}}_{5^2_2}
\ &= \ 
- T_{5^2_2} \int \d^6 \xi \, \e^{-2 \phi} (\det h_{IJ})
\nn \\
\ & \ \ \ \ 
\times
\sqrt{
- \det \Big( 
\wt{g}_{ab} + \frac{\lambda^2 \, \e^{2 \phi} \wt{F}^{(1)}_a \wt{F}^{(1)}_b}{\det h_{IJ}}
\Big)
}
\sqrt{
\det \Big(
\delta_a{}^b 
+ \frac{\i \, \e^{\phi}}{3! \wt{\cal N} \sqrt{(\det h_{IJ}) (\wt{\del a})^2}} \, Z_a{}^b
\Big)
}
\nn \\
\ & \ \ \ \ 
- \frac{\lambda^2}{4} T_{5^2_2} \int \d^6 \xi \, 
\frac{\ve^{abgd'e'f'} \, 
\wt{H}_{d'e'f'} \wt{H}_{abc} (\del_g a \, \del_d a)}{3! \wt{\cal N}^2 (\wt{\del a})^2}
\left[
\wt{g}^{cd} 
- \frac{\lambda^2 \, \e^{2 \phi} \, \wt{g}^{ce} \wt{df} \, \wt{F}^{(1)}_e \wt{F}^{(1)}_f}{\det h_{IJ} + \lambda^2 \, \e^{2 \phi} \, \wt{g}^{a'b'} \wt{F}^{(1)}_{a'} \wt{F}^{(1)}_{b'}}
\right]
\nn \\
\ & \ \ \ \ 
+ T_{5^2_2} \int_{\mathcal{M}_6}
\left(
P [i_{k_1} i_{k_2} B^{(8,2)}]
+ \frac{\lambda^2}{2} \, F^{(3)} \w P [\wt{B}] \w \wt{F}^{(1)}
\right)
\, ,
\label{eq:E8E8522}
\end{align}
where we have defined
\begin{eqnarray}
\begin{aligned}
(\wt{\del a})^2
\ &= \ 
\wt{g}^{ab} \, \del_a a \, \del_b a
\, , \\
\wt{F}_a^{(1)}
\ &= \ 
\del_a Y
\, , \\
Z_a{}^b
\ &= \ 
\frac{\ve^{gbcdef} \big( 
\wt{g}_{ac} 
+ \lambda^2 \, \e^{2 \phi} (\det h_{IJ})^{-1}
\wt{F}^{(1)}_a \wt{F}^{(1)}_b \big) \, \wt{H}_{def} \, \del_g a}{\sqrt{
- \det \big( 
\wt{g}_{a'b'} + \lambda^2 \, \e^{2 \phi} (\det h_{IJ})^{-1} \wt{F}^{(1)}_{a'} \wt{F}^{(1)}_{b'} \big)
}}
\, .
\end{aligned}
\end{eqnarray}
In the string duality chains (see Figure \ref{fig2}),
we can see that the $S^1/{\mathbb Z}_2$ compactification of the M5-brane yields the
$E_8 \times E_8$ heterotic five-brane \cite{Horava:1995qa}. 
The orbifold compactification of the eleven dimensional supergravity 
leaves only the NS-NS sector of type IIA supergravity.
As far as only the Abelian part is concerned,
the ${\cal N}=(2,0)$ tensor multiplet in type IIA theory cannot be distinguished from the combination with ${\cal N}=(1,0)$ tensor multiplet
and hypermultiplet in heterotic string theories. 
Thus the heterotic $5^2_2$-brane effective action can also be given by
the truncation of the R-R sector in the type IIA $5^2_2$-brane effective
action. 
Indeed, when all the R-R potentials are dropped out 
in the type IIA $5^2_2$-brane action \eqref{eq:IIA522}, 
we obtain the action \eqref{eq:E8E8522}.

\section{Conclusion and discussions}
\label{sect-conclusion}

In this paper we studied the world-volume effective actions of the exotic
five-branes in string theories.
Through the duality chains, we explicitly write down the world-volume
effective actions of the $5^2_2$-brane 
and the $5^2_3$-brane 
in type IIB string theory 
and the $5^2_2$-brane in type IIA string theory.
The actions are written in a fully space-time covariant form 
where the two Killing vectors associated with the isometries of the background fields are manifest.

For type IIB case, the world-volume effective action is governed by the
six-dimensional $\mathcal{N} = (1,1)$ vector multiplet. 
The action is obtained by the DBI action of the D5-brane with the Wess-Zumino term
via S- and T-dualities.
The world-volume effective theory consists of four scalar fields $X^1$,
$X^2$, $\varphi$, $\varphi'$ and one $U(1)$ gauge field $\wt{A}_a$. 
The scalar fields $X^1$, $X^2$ are
fluctuation zero-modes along the two transverse directions of the
$5^2_2$-brane and the $5^2_3$-brane. 
The other scalars $\varphi$ and $\varphi'$ are associated with the two transverse
isometry directions (one from the Taub-NUT isometry direction of the
KK-monopole, the other is the additional isometry direction needed for
the second T-duality transformation). 
However, the latter two scalars are not the fluctuation modes since these
directions correspond to the genuine isometries of the (non-)geometry.
For type IIA case, the massless fields of the world-volume effective theory 
belong to the six-dimensional $\mathcal{N} = (2,0)$ tensor multiplet. 
There are five scalar fields $Y$, $X^1$, $X^2$, $\varphi$, $\varphi'$
and one self-dual two-form gauge field $\wt{A}_{ab}$. 
The action is written in the space-time covariant form and also has the
manifestly Lorentz invariant expression with the help of the auxiliary field $a$.
The action is obtained from the PST action of the M5-brane in eleven dimensions 
through the direct dimensional reduction and T-dualities. 
The scalar field $Y$ is originated via the compactification from $D=11$ to $D =10$.
The two scalars
$X^1$, $X^2$ are two transverse fluctuation modes of the IIA
$5^2_2$-brane and two scalars $\varphi$, $\varphi'$ correspond to two
isometry directions. 

The exotic branes are sources of the non-geometric flux 
(the Q-flux \cite{Hassler:2013wsa}, or the mixed-symmetry tensor
\cite{Meessen:1998qm, Bergshoeff:2011zk}).
As discussed in \cite{Chatzistavrakidis:2013jqa}, 
the top couplings of the Wess-Zumino terms of the effective actions are
given by the Q-fluxes $B^{(8,2)}$ and $C^{(8,2)}$.
The gauge transformations of the Q-fluxes are determined such that the
Wess-Zumino term is invariant under the transformations of the NS-NS
$B$-field and the R-R potentials. 

We also discussed the exotic five-branes in type I and heterotic string theories.
Remarkably, the gauge groups of the world-volume effective theories are non-Abelian.
We presented a $U(1)$ sector of the effective actions of the
$5^2_3$-brane in type I and the $5^2_2$-branes in $SO(32)$, $E_8 \times
E_8$ heterotic theories.
Geometrically, this would correspond to the neutral solution of the
five-branes where the $SO(32)$ or $E_8 \times E_8$ gauge field is turned off.
Although we considered only a $U(1)$ part of the world-volume actions in these theories, 
it is important to bear in mind that the non-Abelian gauge fields and
hypermultiplets, that transforms as appropriate representations of the gauge groups, 
need to be incorporated for consistent anomaly free theories.

In this paper, we determined the bosonic part of the world-volume effective actions of the exotic five-branes. 
Since the effective theories in this paper preserve eight or sixteen supercharges, 
the fermion terms are necessary to exhibit the kappa-symmetry.
Although one of the specific properties of the exotic branes
is its global monodromy, the world-volume effective actions studied in
this paper seem not to capture that.
This is because the brane effective actions possess only the local
dynamics (brane fluctuation) near the center of the position of the branes.
In order to explore the global aspects around the exotic branes, 
it would be important to study intersecting configurations of (exotic) branes. 
These are realized as BPS states in the world-volume theory.
Co-dimension two branes (known as defect branes) are not well-defined 
finite energy solutions as the stand alone objects. 
It is also important to study the effective actions including other duality
branes. We will come back to these issues in future researches.

\section*{Acknowledgements}

The authors would like to thank
Masaki Shigemori for discussions.
They would also like to thank Yolanda Lozano for information on 
world-volume theories of exotic branes with two isometries.
They are grateful to the Yukawa Institute for Theoretical Physics at
Kyoto University. Discussions during the YITP molecule-type workshop on
``Exotic Structures of Spacetime'' (YITP-T-13-07) were useful to
complete this work. 
The work of TK is supported in part by the Iwanami-Fujukai Foundation.
The work of SS is supported in part by Kitasato University Research Grant for Young Researchers.

\section*{Appendix}
\begin{appendix}

\section{Conventions and notations of $D = 10$ supergravity}
\label{app-convention}

We use the conventions employed in \cite{Myers:1999ps} where the
gauge transformations of the NS-NS and R-R fields leave the standard Wess-Zumino
term in the D-brane world-volume theory invariant.

We start from the $D=11$ supergravity action:
\begin{align}
S^{(11)} 
\ &= \ 
\frac{1}{2 \kappa^2_{11}} \int \! \d^{11} x \, 
\sqrt{-\wh{g}} \, 
\Big( \wh{R} - \frac{1}{2} |\wh{F}^{(4)}|^2 \Big)
- \frac{1}{12 \kappa^2_{11}} \int \! \wh{C}^{(3)} \wedge \wh{F}^{(4)} \wedge \wh{F}^{(4)}
\, ,
\end{align}
where the fields with the hat symbol stand for the eleven dimensional quantities.
$\kappa_{11}$ is the eleven dimensional gravitational constant, 
$\wh{g}_{MN} \ (M,N = 0, \ldots, 10)$ is the eleven dimensional metric, $\wh{R}$ is the Ricci
scalar, $\wh{C}^{(3)}$ is the three-form potential and $\wh{F}^{(4)} = \d \wh{C}^{(3)}$ is its field strength.
We adopt the convention
$|\wh{F}^{(p)}|^2 = \frac{1}{p!} \wh{F}_{M_1 \cdots M_p} \wh{F}^{M_1 \cdots M_p}$.

The $D=10$ type IIA supergravity action is obtained by the dimensional
reduction of $S^{(11)}$.
The KK ansatz is 
\begin{align}
\wh{g}_{MN} 
\ &= \
\left(
\begin{array}{cc}
\e^{-\frac{2}{3} \phi} \, g_{\mu \nu} 
+ \e^{\frac{4}{3} \phi} \, C^{(1)}_{\mu} C^{(1)}_{\nu} 
& \e^{\frac{4}{3} \phi} \, C^{(1)}_{\mu} 
\\
\e^{\frac{4}{3} \phi} \, C^{(1)}_{\nu} 
& \e^{\frac{4}{3} \phi}
\end{array}
\right)
\, ,
\end{align}
where $\mu, \nu = 0, \ldots, 9$ are space-time indices in ten dimensions.
The fields $g_{\mu \nu}$, $\phi$ and $C^{(1)}_{\mu}$ are the space-time
metric, the dilaton, and the R-R one-form potential in ten dimensions,
respectively.
The ten-dimensional type IIA supergravity action is given by 
\begin{align}
S_{\text{IIA}} 
\ &= \ 
\frac{1}{2\kappa^2_{10}} \int \! \d^{10} x \, \sqrt{-g} \, \e^{-2\phi} 
\left(
R + 4 (\partial_{\mu} \phi)^2 - \frac{1}{2} |H^{(3)}|^2
\right)
\nn \\
\ & \ \ \ \ 
- \frac{1}{4 \kappa^2_{10}} \int \! \d^{10} x \, \sqrt{-g} \,
\left(
|G^{(2)}|^2 + |G^{(4)}|^2
\right)
- \frac{1}{4 \kappa^2_{10}} \int \! B \wedge F^{(4)} \wedge F^{(4)}
\, ,
\label{eq:IIAsugra}
\end{align}
where the modified field strengths are defined by 
\begin{eqnarray}
\begin{gathered}
F^{(2)} \ = \ \d C^{(1)}
\, , \ls
H^{(3)} \ = \ \d B
\, , \ls
F^{(4)} \ = \ \d C^{(3)}
\, , \\
G^{(2)} \ = \ F^{(2)}
\, , \ls
G^{(4)} \ = \ 
F^{(4)} + H^{(3)} \wedge C^{(1)}
\, .
\end{gathered}
\end{eqnarray}
Here $C^{(n)}$ is the R-R $n$-form 
potential and $F^{(n+1)} = \d C^{(n)}$ is its field strength.
$H^{(3)} = \d B$ is the field strength of the NS-NS $B$-field, and
$R$ is the Ricci scalar in ten dimensions.
The equations of motion for the R-R fields are 
\begin{eqnarray}
\begin{aligned}
\delta C^{(1)} \ &: &\ \ \
0 \ &= \ 
\d (*_{10} G^{(2)}) - H^{(3)} \wedge *_{10} G^{(4)} 
\, , \\
\delta C^{(3)} \ &: &\ \ \
0 \ &= \ 
\d (*_{10} G^{(4)}) - H^{(3)} \wedge G^{(4)}
\, .
\end{aligned}
\label{eq:IIARReom}
\end{eqnarray}
Here $*_{10}$ is the Hodge dual operator in ten dimensions.
The Bianchi identities are 
\begin{gather}
\d H^{(3)} \ = \ \d F^{(2)} \ = \ 0
\, , \ls
\d G^{(4)} \ = \ - H^{(3)} \wedge G^{(2)}
\, .
\end{gather}
We define the dual field strengths of the R-R fields as 
\begin{eqnarray}
\begin{aligned}
G^{(8)} \ &= \ *_{10} G^{(2)}
\, , &\ls
G^{(6)} \ &= \ - *_{10} G^{(4)}
\, .
\end{aligned}
\end{eqnarray}
The equations of motion \eqref{eq:IIARReom} become the Bianchi
identities for the dual fields, 
\begin{gather}
\d G^{(6)} 
\ = \ - H^{(3)} \wedge G^{(4)}
\, , \ls
\d G^{(8)} 
\ = \ - H^{(3)} \wedge G^{(6)}
\, .
\label{eq:IIARRbianchi}
\end{gather}
The modified field strengths that satisfy the Bianchi identities
\eqref{eq:IIARRbianchi} are given by 
\begin{gather}
G^{(6)} \ = \ F^{(6)} + H^{(3)} \wedge C^{(3)}
\, , \ls
G^{(8)} \ = \ F^{(8)} + H^{(3)} \wedge C^{(5)}
\, .
\end{gather}
The equation of motion for $B$ is 
\begin{align}
\delta B \ : \ \ \
0 \ &= \ 
\d (\e^{-2\phi} *_{10} H^{(3)}) 
- G^{(2)} \wedge *_{10} G^{(4)} 
- \frac{1}{2} G^{(4)} \wedge G^{(4)} 
\, . 
\label{eq:IIABeom}
\end{align}
We define the dual field strength of the NS-NS $B$-field as 
\begin{align}
\check{H}^{(7)} \ = \ \e^{-2\phi} *_{10} H^{(3)}
\, .
\end{align}
Then the equation of motion \eqref{eq:IIABeom} becomes the 
Bianchi identity for $\check{H}^{(7)}$:
\begin{align}
 \d \check{H}^{(7)} 
+ G^{(2)} \wedge G^{(6)} 
- \frac{1}{2} G^{(4)} \wedge G^{(4)} 
\ &= \ 0 
\, .
\label{eq:H7IIA-Bianchi}
\end{align}
Therefore the dual field strength $\check{H}^{(7)}$ that satisfies the
Bianchi identity \eqref{eq:H7IIA-Bianchi} is given by 
\begin{align}
\check{H}^{(7)} 
\ &= \ 
\d B^{(6)} 
- \frac{1}{2} G^{(2)} \wedge C^{(5)} 
+ \frac{1}{2} G^{(4)} \wedge C^{(3)} 
- \frac{1}{2} G^{(6)} \wedge C^{(1)}
\, .
\end{align}
The gauge transformations are defined such that the modified field strengths are invariant. 
The explicit expressions of the transformation rules are found to be as follows:
\begin{eqnarray}
\begin{aligned}
\delta B \ &= \ 
\d \Lambda^{(1)}
\, , \\
\delta B^{(6)} \ &= \ 
\d \Lambda^{(5)} 
+ \frac{1}{2} C^{(1)} \wedge \d \lambda^{(4)}
- \frac{1}{2} \left( C^{(3)} + C^{(1)} \wedge B \right) \wedge \d \lambda^{(2)} 
\\
\ & \ \ \ \  
+ \frac{1}{2} 
\left(
C^{(5)} + C^{(3)} \wedge B + \frac{1}{2} C^{(1)} \wedge B \wedge B
\right) \wedge \d \lambda^{(0)}
\, , \\
\delta C^{(1)} \ &= \ 
\d \lambda^{(0)}
\, , \\
\delta C^{(3)} \ &= \ 
\d \lambda^{(2)} - B \wedge \d \lambda^{(0)}
\, , \\
\delta C^{(5)} \ &= \ 
\d \lambda^{(4)} 
- B \wedge \d \lambda^{(2)} 
+ \frac{1}{2!} B \wedge B \wedge \d \lambda^{(0)}
\, , \\
\delta C^{(7)} \ &= \ 
\d \lambda^{(6)} 
- B \wedge \d \lambda^{(4)} 
+ \frac{1}{2!} B \wedge B \wedge \d \lambda^{(2)}
- \frac{1}{3!} B \wedge B \wedge B \wedge \d \lambda^{(0)}
\, .
\end{aligned}
\end{eqnarray}
Here $\Lambda^{(1)}, \Lambda^{(5)}, \lambda^{(n)} \ (n=0,2,4,6)$ are
gauge parameters.
The transformation rules of the R-R fields are summarized as
\begin{align}
\delta C^{(p)} 
\ &= \ 
\sum_p \d \lambda^{(n)} \wedge \e^{-B}
\, ,
\end{align}
which leave the Wess-Zumino term in the D-brane world-volume theory invariant.

The ten-dimensional type IIB supergravity action is given by 
\begin{align}
S_{\mathrm{IIB}} 
\ &= \ 
\frac{1}{2\kappa^2_{10}} \int \! \d^{10} x \, \sqrt{-g} \, \e^{-2\phi}
\left(
R + 4 (\partial_{\mu} \phi)^2 - \frac{1}{2} |H^{(3)}|^2
\right)
\nn \\
\ & \ \ \ \ 
- \frac{1}{4\kappa^2_{10}} \int \! \d^{10} x \, \sqrt{-g} \,
\left(
|G^{(1)}|^2 + |G^{(3)}|^2 + \frac{1}{2} |G^{(5)}|^2
\right)
\nn \\
\ & \ \ \ \ 
+ \frac{1}{4 \kappa^2_{10}} \int \! 
\left(
C^{(4)} + \frac{1}{2} B \wedge C^{(2)}
\right) \wedge H^{(3)} \wedge F^{(3)}
\, .
\label{eq:IIBsugra}
\end{align}
Note that,
in contrast to the ones in \cite{Myers:1999ps}, we change the sign in front of the Chern-Simons term.
The modified field strengths are given by 
\begin{align}
G^{(1)} \ = \ F^{(1)}
\, , \ls
G^{(3)} 
\ = \ 
F^{(3)} + H^{(3)} C^{(0)}
\, , \ls
G^{(5)} 
\ = \ F^{(5)} + H^{(3)} \wedge C^{(2)}
\, .
\end{align}
In addition, at the level of the equation of motion, 
we impose the self-duality condition on $G^{(5)}$:
\begin{align}
G^{(5)} \ = \ *_{10} G^{(5)}
\, .
\label{eq:sd}
\end{align}
The equations of motion for the R-R fields are 
\begin{eqnarray}
\begin{aligned}
\delta C^{(0)} \ &: &\ \ \ 
0 \ &= \ 
\d (*_{10} G^{(1)}) + *_{10} G^{(3)} \wedge H^{(3)}
\, , \\
\delta C^{(2)} \ &: &\ \ \ 
0 \ &= \ 
\d (*_{10} G^{(3)}) 
+ *_{10} G^{(5)} \wedge H^{(3)}
\, , \\
\delta C^{(4)} \ &: &\ \ \ 
0 \ &= \ 
\d (*_{10} G^{(5)}) + H^{(3)} \wedge G^{(3)} 
\, .
\end{aligned}
\end{eqnarray}
The last equation is automatically satisfied when the self-duality
condition \eqref{eq:sd} is imposed.
The Bianchi identities are
\begin{eqnarray}
\begin{aligned}
\d H^{(3)} \ &= \ 0
\, , \ls
\d G^{(1)} \ = \ 0
\, , \ls
\d G^{(3)} \ = \ - H^{(3)} \wedge G^{(1)}
\, , \ls
\d G^{(5)} \ = \ - H^{(3)} \wedge G^{(3)}
\, .
\end{aligned}
\end{eqnarray}
The dual fields are defined as
\begin{eqnarray}
\begin{aligned}
G^{(7)} \ &= \ - *_{10} G^{(3)}
\, , &\ls
G^{(9)} \ &= \ *_{10} G^{(1)}
\, .
\end{aligned}
\end{eqnarray}
The Bianchi identities for the dual fields are 
\begin{eqnarray}
\begin{aligned}
\d G^{(9)} \ &= \ - H^{(3)} \wedge G^{(7)}
\, , \ls
\d G^{(7)} \ = \ - H^{(3)} \wedge G^{(5)}
\, .
\end{aligned}
\end{eqnarray}
The modified field strengths of the dual fields that satisfy the Bianchi
identities are found to be 
\begin{eqnarray}
\begin{aligned}
G^{(7)} \ &= \ \d C^{(6)} + H^{(3)} \wedge C^{(4)}
\, , \\
G^{(9)} \ &= \ \d C^{(8)} + H^{(3)} \wedge C^{(6)}
\, .
\end{aligned}
\end{eqnarray}
The equation of motion for the NS-NS $B$-field is 
\begin{align}
0 \ &= \ 
\d (\e^{-2\phi} *_{10} H^{(3)}) 
+ G^{(1)} \wedge *_{10} G^{(3)} 
+ G^{(3)} \wedge *_{10} G^{(5)} 
\, .
\end{align}
This gives the Bianchi identity for the dual field:
\begin{align}
\d \check{H}^{(7)} 
- G^{(1)} \wedge G^{(7)} 
+ G^{(3)} \wedge G^{(5)} 
\ &= \ 0
\, .
\end{align}
Then the dual field strength is given by 
\begin{align}
\check{H}^{(7)} 
\ &= \ 
\d B^{(6)} 
+ C^{(4)} \wedge F^{(3)} 
- \frac{1}{2} C^{(2)} \wedge C^{(2)} \wedge H^{(3)} 
+ C^{(0)} G^{(7)}
\, .
\end{align}
The gauge transformations are defined such that the modified field
strengths are invariant:
\begin{eqnarray}
\begin{aligned}
\delta B \ &= \ \d \Lambda^{(1)}
\, , \\
\delta B^{(6)} \ &= \ 
\d \Lambda^{(5)} 
- \d \lambda^{(3)} \wedge C^{(2)} 
+ \d \lambda^{(1)} \wedge B \wedge C^{(2)}
\, , \\
\delta C^{(0)} \ &= \ 0
\, , \\
\delta C^{(2)} \ &= \ 
\d \lambda^{(1)}
\, , \\
\delta C^{(4)} \ &= \ 
\d \lambda^{(3)} - B \wedge \d \lambda^{(1)}
\, , \\
\delta C^{(6)} \ &= \ 
\d \lambda^{(5)} 
- B \wedge \d \lambda^{(3)} 
+ \frac{1}{2} B \wedge B \wedge \d \lambda^{(1)}
\, , \\
\delta C^{(8)} \ &= \ 
\d \lambda^{(7)} 
- B \wedge \d \lambda^{(5)} 
+ \frac{1}{2} B \wedge B \wedge \d \lambda^{(3)} 
- \frac{1}{3!} B \wedge B \wedge B \wedge \d \lambda^{(1)}
\, .
\end{aligned}
\end{eqnarray}
Here $\lambda^{(n)} \ (n=1,3,5,7)$ are gauge parameter $n$-forms.
Hence the gauge transformations of the R-R fields are 
\begin{align}
\delta C^{(p)} \ &= \
\sum_p \d \lambda^{(n)} \wedge \e^{-B}
\, .
\end{align}
Again, these gauge transformations leave the Wess-Zumino term in the
D-brane world-volume action invariant.
The type IIB supergravity has the $SL(2, \mathbb{R})$ symmetry:
\begin{eqnarray}
\begin{aligned}
g_{\mu \nu} \ &\to \ 
|\tau| \, g_{\mu \nu}
\, , \\
\tau \ &\to \ - \frac{1}{\tau}
\, , \ls
C^{(0)} \ \to \ 
- \frac{C^{(0)}}{|\tau|^2}
\, , \ls
\e^{-\phi} \ \to \ 
\frac{e^{-\phi}}{|\tau|^2}
\, , \\
F_3^i \ &\to \ 
\Lambda^i {}_j F_3^j
\, , \ls
\Lambda^i {}_j \ = \ 
\left(
\begin{array}{cc}
0 & -1 \\
1 & 0
\end{array}
\right) \in SL(2,\mathbb{R})
\, , \\
C^{(2)} \ &\to \ B
\, , \ls
B \ \to \ - C^{(2)}
\, , \ls
C^{(4)} \ \to \ C^{(4)} + C^{(2)} \wedge B
\, ,
\end{aligned}
\end{eqnarray}
where we have defined 
\begin{eqnarray}
\begin{aligned}
\tau \ &= \ 
C^{(0)} + \i \e^{-\phi}
\, , \ls
|\tau|^2 \ = \ (C^{(0)})^2 + \e^{-2\phi}
\, , \\
M_{ij} \ &= \ 
\frac{1}{\Im \tau}
\left(
\begin{array}{cc}
|\tau|^2 & - \Re \tau \\
- \Re \tau & 1 
\end{array}
\right) 
\ = \ \e^{\phi} 
\left(
\begin{array}{cc}
(C^{(0)})^2 + \e^{-2\phi} & - C^{(0)} \\
- C^{(0)} & 1
\end{array}
\right)
\, , \\
F_3^i \ &= \ 
\left(
\begin{array}{c}
H^{(3)} \\
F^{(3)}
\end{array}
\right)
\, .
\end{aligned}
\end{eqnarray}
This is nothing but the S-duality symmetry.
The S-duality transformations of the dual fields are
\begin{eqnarray}
\begin{aligned}
C^{(6)} \ &\to \ - B^{(6)} + \frac{1}{2} B \wedge C^{(2)} \wedge C^{(2)}
\, , \\
B^{(6)} \ &\to \ C^{(6)} - \frac{1}{2} C^{(2)} \wedge B \wedge B
\, .
\end{aligned}
\end{eqnarray}
Under the S-duality, D5-branes transform to NS5-branes. 
One can confirm that the transformation rules given above
provide the correct gauge invariant Wess-Zumino term of the NS5-brane in
the world-volume effective action.

\section{Covariant T-duality transformation rules for supergravity fields}
\label{app-covBuscher}

In this appendix, we provide the space-time covariant expressions of the
T-duality Buscher rule \cite{Buscher:1987sk} for the NS-NS and R-R
fields.

\subsection{NS-NS sector}

The string world-sheet sigma model action is given by 
\begin{align}
S \ &= \ 
- \frac{1}{2 \pi \alpha'} 
\int \! \d^2 \sigma \, \sqrt{-\gamma} \,
\left[
(\gamma^{ab} g_{\mu \nu} + \epsilon^{ab} B_{\mu \nu}) \partial_a X^{\mu}
 \partial_b X^{\nu}
\right]
\, ,
\end{align}
where $\gamma_{ab}$ is the world-sheet metric, 
while $g_{\mu \nu}$, $B_{\mu\nu}$ are the background space-time metric and 
the NS-NS $B$-field.
We have omitted the coupling to the dilaton term for simplicity.
The background fields $g_{\mu \nu}$ and
$B_{\mu \nu}$ have isometries 
$X^{\mu} \to X^{\mu} + \eta^I k^{\mu}_I \ (I=1, \ldots, N)$ 
generated by the Killing vectors $k^{\mu}_I$.
Here $\eta^I$ are the transformation parameters.
Now we gauge the isometry on the world-sheet.
Then the world-sheet action becomes
\begin{align}
S_{\text{gauged}} 
\ &= \
- \frac{1}{2 \pi \alpha'} \int \! \d^2 \sigma \, \sqrt{-\gamma} \,
\left(
\gamma^{ab} g_{\mu \nu} + \epsilon^{ab} B_{\mu \nu} 
\right) D_a X^{\mu} D_b X^{\nu}
+ \frac{1}{2\pi \alpha'} \int \! \d^2 \sigma \, \lambda \, \varepsilon^{ab}
 \varphi_I F^I_{ab}
\, ,
\end{align}
where $D_a X^{\mu} = \partial_a X^{\mu} + C_a^I k^{\mu}_I$ is the gauge
covariant derivative and 
$\varepsilon^{ab} = \sqrt{-\gamma} \epsilon^{ab}$ is the Levi-Civita antisymmetric symbol.
Here we introduced Lagrange multipliers $\varphi^I$ to ensure that the
auxiliary gauge field $C^I_a$ takes valued in the trivial homotopy class 
$F^I_{ab} = \partial_{a}^{\vphantom{I}} C^I_b - \partial_b^{\vphantom{I}} C^I_a = 0$.
The equation of motion for $C^I_a$ is 
\begin{align}
\Big( \gamma^{ab} \, k^{\mu}_I \, g_{\mu \nu} + \epsilon^{ab} \, k^{\mu}_I B_{\mu \nu} \Big)
 k^{\nu}_J \, C_b^J 
\ &= \ 
- \Big( \gamma^{ab} \, k^{\mu}_I \, g_{\mu \nu} + \epsilon^{ab} \, k^{\mu}_I B_{\mu \nu} \Big)
 \partial_b X^{\nu} + \lambda \, \epsilon^{ab} \partial_b \varphi
\, .
\end{align}
We consider the case where one Killing vector $k^{\mu}$ exists. 
Then the solution is 
\begin{align}
C_a \ &= \ 
- \frac{1}{k^2} 
\left(
\delta_a {}^b (i_k g)_{\mu} 
+ \gamma_{ac} \, \epsilon^{cb} (i_k B - \lambda \d \varphi)_{\mu}
\right) \partial_b X^{\mu}
\, ,
\end{align}
where
\begin{align}
k^2 \ &= \ 
g_{\mu \nu} \, k^{\mu} k^{\nu}
\, , \ls
(i_k g)_{\mu} \ = \ k^{\rho} g_{\rho \mu}
\, , \ls
(i_k B)_{\mu} \ = \ k^{\rho} B_{\rho \mu}
\, .
\end{align}
Plugging the solution back into the action, we find 
\begin{align}
S_{\text{gauged}} 
\ &= \ 
- \frac{1}{2\pi \alpha'} \int \! \d^2 \sigma \,
 \Scr{L}_{\text{gauged}}
\, ,
\end{align}
where 
\begin{align}
\Scr{L}_{\text{gauged}} 
\ &= \
 \sqrt{-\gamma} \, \gamma^{ab}
\left\{
g_{\mu \nu} - \frac{(i_k g)_{\mu} (i_k g)_{\nu} - (i_k B - \lambda \d
 \varphi)_{\mu} (i_k B - \lambda \d \varphi)_{\nu}}{k^2}
\right\} \partial_a X^{\mu} \partial_b X^{\nu}
\nn \\
\ & \ \ \ \ 
+ \sqrt{-\gamma} \, \epsilon^{ab} 
\left\{
B_{\mu \nu} - \frac{(i_k g)_{\mu} (i_k B - \lambda \d \varphi)_{\nu} -
 (i_k B - \lambda \d \varphi)_{\mu} (i_k g)_{\nu}}{k^2}
\right\} \partial_a X^{\mu} \partial_b X^{\nu}
\, .
\end{align}
The T-duality transformation rule of the dilaton is determined at the one-loop level
\cite{Buscher:1987sk}. Making it covariant form, we obtain
the following covariant Buscher rules:
\begin{eqnarray}
\begin{aligned}
g'_{\mu \nu} 
\ &= \ 
g_{\mu \nu} 
- \frac{(i_k g)_{\mu} (i_k g)_{\nu} 
- ( i_k B - \lambda \d \varphi)_{\mu} (i_k B - \lambda \d \varphi)_{\nu}}{k^2}
\, , \\
B'_{\mu \nu} \ &= \ 
B_{\mu \nu} - \frac{(i_k g)_{\mu} 
(i_k B - \lambda \d \varphi)_{\nu} - (i_k B - \lambda \d \varphi)_{\mu} (i_k g)_{\nu}}{k^2}
\, , \\
\e^{2\phi'} \ &= \ 
\frac{1}{k^2} \, \e^{2\phi}
\, .
\end{aligned}
\end{eqnarray}
For the NS-NS $B$ field, the transformation rule is written as 
\begin{align}
B' \ &= \ B - \frac{1}{k^2} (i_k g) \wedge (i_k B - \lambda \d \varphi)
\, , 
\end{align}
where $i_k g$ is treated as a one-form.
We note that the field $\varphi$ is associated with the dual coordinate.
Then we find $\lambda k^{\mu} \partial_{\mu} \varphi = 1$.
In an adapted coordinate $k^{\mu} = \delta^{\mu y}$, we recover the
well-known non-covariant Buscher rule.

The repeated application of the Buscher rule gives 
\begin{eqnarray}
\begin{aligned}
i_{k_1} g \ &\xrightarrow[k_2]{} \ 
i_{k_1} g - \frac{1}{(k_2)^2} 
\left\{
(k_1 \cdot k_2) i_{k_2} g
- (i_{k_1} i_{k_2} B) 
(i_{k_2} B - \lambda \d \varphi')
\right\}
\, , \\
i_{k_1} B \ &\xrightarrow[k_2]{} \ 
i_{k_1} B - \frac{1}{(k_2)^2} \, i_{k_1} 
\left(
i_{k_2} g \wedge (i_{k_2} B - \lambda \d \varphi')
\right)
\, , \\
k_1^2 \ &\xrightarrow[k_2]{} \ 
\frac{\det h_{IJ}}{(k_2)^2} 
\, , \ls 
h_{IJ} \ = \ k_I^{\mu} k_J^{\nu} (g_{\mu \nu} + B_{\mu \nu})
\, , \ \ \ \ (I,J=1,2)
\, .
\end{aligned}
\end{eqnarray}
Here the arrow
``$\xrightarrow[k_2]{}$'' indicates the T-duality transformation along
the isometry generated by the Killing vector $k^{\mu}_2$. 
Performing the T-duality transformations along the two isometry directions,
we obtain $B \xrightarrow[k_1 k_2]{} \wt{B}$ where 
\begin{align}
\wt{B} \ &= \ 
B - \frac{1}{(k_2)^2} (i_{k_2} g) \wedge K^{(1)} 
- \frac{(k_2)^2}{\det h_{IJ}} \, K^{(2)} \wedge K^{(3)}
\, .
\end{align}
Here $K^{(1)}$, $K^{(2)}$, $K^{(3)}$ are given in \eqref{eq:K_def}.

\subsection{R-R sector}

For the R-R sector, the T-duality transformation rules
of the R-R fields are given in the non-covariant expressions in
\cite{Meessen:1998qm}.
We decompose the space-time indices $\mu$ into $y$ (i.e., the T-dual direction), and $\hat{\mu}, \hat{\nu}, \ldots \neq y$.
Then the transformations of the R-R fields are 
\begin{eqnarray}
\begin{aligned}
C'{}^{(n)}_{\hat{\mu}_1 \cdots \hat{\mu}_{n-2} \hat{\nu} y} 
\ &= \ 
C^{(n-1)}_{\hat{\mu}_1 \cdots \hat{\mu}_2 \hat{\nu}} 
- \frac{n-1}{g_{yy}} \,
 C^{(n-1)}_{[\hat{\mu}_1 \cdots \hat{\mu}_{n-2} | y} g_{|\hat{\nu}] y}^{\vphantom{(n-1)}}
\, , \\
C'{}^{(n)}_{\hat{\mu}_1 \cdots \hat{\mu}_{n-2} \hat{\nu} \hat{\rho}} 
\ &= \ 
C^{(n+1)}_{\hat{\mu}_1 \cdots \hat{\mu}_{n-2} \hat{\nu} \hat{\rho} y} 
+ n \, C^{(n-1)}_{[\hat{\mu}_1 \cdots \hat{\mu}_{n-2} \hat{\nu}} 
B_{\hat{\rho}] y}^{\vphantom{(n-1)}}
+ \frac{n(n-1)}{g_{yy}} \,
C^{(n-1)}_{[\hat{\mu}_1 \cdots \hat{\mu}_{n-2} | y} B_{|\hat{\nu}| y}^{\vphantom{(n-1)}} \,
g_{|\hat{\rho}] y}^{\vphantom{(n-1)}}
\, ,
\end{aligned}
\label{eq:RR_T-dual}
\end{eqnarray}
where the antisymmetric symbol is defined such that 
$A_{[\mu_1 \cdots \mu_n]} = \frac{1}{n!} (A_{\mu_1 \cdots
\mu_n} \pm (\mathrm{perm}))$.
Here (perm) stands for terms with possible permutation of indices.
Although these are not covariant forms,
the covariant expression of the T-duality transformation 
for the pull-back of the R-R field is obtained as follows.
The pull-back of the R-R field $C^{(n)}_{\mu_1 \cdots \mu_n}$ is 
\begin{align}
P[C^{(n)}]_{a_1 \cdots a_n} 
\ &= \
n! C^{(n)}_{y \hat{\mu}_2 \cdots \hat{\mu}_n} 
\partial_{[a_1} Y \partial_{a_2} X^{\hat{\mu}_2} \cdots \partial_{a_n]}
 X^{\hat{\mu}_n} 
+ C^{(n)}_{\hat{\mu}_1 \cdots \hat{\mu}_n} \partial_{a_1}
 X^{\hat{\mu}_1} \cdots \partial_{a_n} X^{\hat{\mu}_n}.
\end{align}
Here $Y$ is the world-volume scalar field associated with the T-dual
direction $y$. 
Using the T-duality rule \eqref{eq:RR_T-dual}, $P[C^{(n)}]$
transforms as
\begin{align}
P[C^{(n) \prime}]_{a_1 \cdots a_n} \ &= \ 
 n! (-)^{n-1} C^{(n-1)}_{\hat{\mu}_2 \cdots \hat{\mu}_{n}} 
\partial_{[a_1} Y' \cdots \partial_{a_n]} X^{\hat{\mu}_n} 
\nn \\
\ & \ \ \ \ 
+ n! (-)^{n} (i_k C^{(n+1)})_{\hat{\mu}_1 \cdots \hat{\mu}_{n}}
\partial_{[a_1 } X^{\hat{\mu}_1} \cdots \partial_{a_n]} 
X^{\hat{\mu}_n} 
\nn \\
\ & \ \ \ \ 
- n! \, n \, C^{(n-1)}_{[ \hat{\mu}_1 \cdots \hat{\mu}_{n-1}} (i_k B)_{\hat{\mu}_n]}^{\vphantom{(n-1)}} 
\partial_{[a_1}^{\vphantom{(n-1)}}  X^{\hat{\mu}_1} \cdots \partial_{a_n]}^{\vphantom{(n-1)}}  X^{\hat{\mu}_n} 
\nn \\
\ & \ \ \ \  
+ n! \frac{n-1}{k^2} (i_k C^{(n-1)})_{[\hat{\mu}_2 \cdots
 \hat{\mu}_{n-1}} (i_k g)_{\hat{\mu}_n]}^{\vphantom{(n-1)}} \partial_{[a_1}^{\vphantom{(n-1)}} Y' \cdots
 \partial_{a_n]}^{\vphantom{(n-1)}} X^{\hat{\mu}_n} 
\nn \\
\ & \ \ \ \ 
- n! (-)^{n-2} \frac{n(n-1)}{k^2} (i_k C^{(n-1)})_{[\hat{\mu}_1 \cdots
 \hat{\mu}_{n-2}} (i_k B)_{\hat{\mu}_{n-1}} (i_k g)_{\hat{\mu}_n]} 
\partial_{[a_1} X^{\hat{\mu}_1} \cdots \partial_{a_n]} X^{\hat{\mu}_n}
\, ,
\label{eq:RR_T-dual_cov}
\end{align}
where we have used the fact that $g_{yy} = k^2$, $C^{(n)}_{\hat{\mu}_1 \cdots
\hat{\mu}_{n-1} y} = (i_{k} C^{(n)})_{\hat{\mu}_1 \cdots \hat{\mu}_{n-1}}$ and so on.
Here we defined the dual coordinate of $Y$ as $Y'$
and $i_k C^{(n)}$ is the interior product of $k^{\mu}$ and $C^{(n)}$.
Due to the antisymmetric nature of indices, we can replace $\hat{\mu}$
with $\mu$ in the first, second and the fourth lines in
\eqref{eq:RR_T-dual_cov}.
Careful treatments of terms in the combination of the third and fifth
lines in \eqref{eq:RR_T-dual_cov} allow one to replace the indices
$\hat{\mu}$ with $\mu$ in these terms.
Collecting all together, we find the space-time covariant Buscher rule
for the pull-back of R-R fields as follows:
\begin{align}
P[C^{(n) \prime}]_{a_1 \cdots a_n} 
\ &= \ 
\left[
(-)^n (i_k C^{(n+1)})_{\mu_1 \cdots \mu_n}^{\vphantom{(n-1)}}  
- n \, C^{(n-1)}_{[\mu_1 \cdots \mu_{n-1}} 
(i_k B - \lambda \d \varphi)_{\mu_n ]}^{\vphantom{(n-1)}}
\right. 
\nn \\
\ & \ \ \ \ 
\left. 
- (-)^{n-2} \frac{n(n-1)}{k^2} (i_k C^{(n-1)})_{[\mu_1 \cdots \mu_{n-2} }
 (i_k B - \lambda \d \varphi )_{\mu_{n-1}} (i_k g)_{\mu_{n} ]} 
\right] \partial_{a_1} X^{\mu_1} \cdots \partial_{a_n} X^{\mu_n}
\, ,
\end{align}
where we have redefined 
$Y' = \lambda \varphi$.
Therefore, the R-R field in the pull-back is transformed as
\begin{align}
C^{(n) \prime} 
\ &= \ 
(-)^n i_k C^{(n+1)} 
- C^{(n-1)} \wedge (i_k B - \lambda \d \varphi) 
- \frac{(-)^{n-2}}{k^2} \, i_k C^{(n-1)} \wedge (i_k B - \lambda \d \varphi) \wedge i_k g
\, .
\end{align}

Performing the T-duality transformations along the two isometry directions,
the R-R potentials become
\bsubeq
\begin{align}
\wt{C}^{(0)} \ &= \ - i_{k_1} i_{k_2} (C^{(2)} + C^{(0)} B),
\nn \\
\wt{C}^{(1)} 
\ &= \  
- i_{k_1} i_{k_2} C^{(3)} 
+ (i_{k_1} C^{(1)}) K^{(1)} 
- (i_{k_1} i_{k_2} B) C^{(1)} 
+ \frac{1}{(k_2)^2} (i_{k_2} C^{(1)}) (i_{k_1} i_{k_2} B) \, (i_{k_2} g) 
\nn \\
\ & \ \ \ \ 
- \frac{1}{(k_2)^2} (i_{k_2} C^{(1)}) (k_1 \cdot k_2) K^{(1)}
- (i_{k_2} C^{(1)}) K^{(3)}
\, , \\
\wt{C}^{(2)} 
\ &= \ 
- i_{k_1} i_{k_2} C^{(4)} 
- (i_{k_1} C^{(2)}) \wedge K^{(1)} 
- (i_{k_1} i_{k_2} B) C^{(2)} 
\nn \\
\ & \ \ \ \ 
+ \frac{1}{(k_2)^2} (i_{k_1} i_{k_2} C^{(2)}) K^{(1)} \wedge (i_{k_2} g) 
- \frac{1}{(k_2)^2} (i_{k_1} i_{k_2} B) \, (i_{k_2} C^{(2)}) \wedge (i_{k_2} g)
\nn \\
\ & \ \ \ \ 
+ \frac{k_1 \cdot k_2}{(k_2)^2} (i_{k_2} C^{(2)}) \wedge K^{(1)}
+ \left[
i_{k_2} C^{(2)} + C^{(0)} K^{(1)}
\right]
\wedge K^{(3)} 
\nn \\
\ & \ \ \ \  
+ \frac{(k_2)^2}{\det h_{IJ}} \Big[ i_{k_1} i_{k_2} (C^{(2)} + C^{(0)} B) \Big]
 K^{(3)} \wedge K^{(2)}
\, , \\
\wt{C}^{(3)}
\ &=  \ 
- i_{k_1} i_{k_2} C^{(5)} 
+ (i_{k_1} C^{(3)}) \wedge K^{(3)} 
- (i_{k_1} i_{k_2} B) C^{(3)} 
\nn \\
\ & \ \ \ \ 
+ \frac{1}{(k_2)^2} (i_{k_1} i_{k_2} C^{(3)}) \wedge K^{(1)} \wedge (i_{k_2} g) 
+ \frac{1}{(k_2)^2} (i_{k_1} i_{k_2} B) (i_{k_2} C^{(3)}) \wedge (i_{k_2} g) 
- \frac{k_1 \cdot k_2}{(k_2)^2} (i_{k_2} C^{(3)}) \wedge K^{(1)} 
\nn \\
\ & \ \ \ \  
- \left[
i_{k_2} C^{(3)} - C^{(1)} \wedge K^{(1)} 
- \frac{1}{(k_2)^2} (i_{k_2} C^{(1)})
 K^{(1)} \wedge (i_{k_2} g)
\right] \wedge K^{(3)}
\nn \\
\ & \ \ \ \ 
+ \frac{(k_2)^2}{\det h_{IJ}} 
\left[
i_{k_1} i_{k_2}  C^{(3)} 
- (i_{k_2} C^{(1)}) K^{(1)} 
+ (i_{k_1} i_{k_2} B) C^{(1)} 
- \frac{1}{(k_2)^2} (i_{k_2} C^{(1)}) (i_{k_1} i_{k_2} B) (i_{k_2} g) 
\right. 
\nn \\
\ & \LS \LS 
\left. 
+ \frac{1}{(k_2)^2} (i_{k_2} C^{(1)}) (k_1 \cdot k_2) K^{(1)}
\right]
\wedge K^{(3)} \wedge K^{(2)}
\, ,
\end{align}
\esubeq
and 
\bsubeq
\begin{align}
\wt{C}^{(4)} 
\ &= \ 
- i_{k_1} i_{k_2} C^{(6)} 
- (i_{k_1} C^{(4)}) \wedge K^{(1)} 
- (i_{k_1} i_{k_2} B ) C^{(4)} 
\nn \\
\ & \ \ \ \ 
+ \frac{1}{(k_2)^2} (i_{k_1} i_{k_2} C^{(4)}) \wedge K^{(1)} \wedge (i_{k_2} g) 
- \frac{1}{(k_2)^2} (i_{k_1} i_{k_2} B) (i_{k_2} C^{(4)}) \wedge (i_{k_2} g) 
+ \frac{k_1 \cdot k_2}{(k_2)^2} (i_{k_2} C^{(4)}) \wedge K^{(1)} 
\nn \\
\ & \ \ \ \ 
+ \left[
i_{k_2} C^{(4)} 
+ C^{(2)} \wedge K^{(1)} 
- \frac{1}{(k_2)^2} (i_{k_2} C^{(2)}) \wedge K^{(1)} \wedge (i_{k_2} g)
\right] \wedge K^{(3)} 
\nn \\
\ & \ \ \ \ 
+ \frac{(k_2)^2}{\det h_{IJ}} 
\left\{
i_{k_1} i_{k_2} C^{(4)} 
+ (i_{k_1} C^{(2)}) \wedge K^{(1)} + (i_{k_1} i_{k_2} B) C^{(2)}
- \frac{1}{(k_2)^2} (i_{k_1} i_{k_2} C^{(2)}) K^{(1)} \wedge i_{k_2} g
\right.
\nn \\
\ & \LS \LS
\left.
+ \frac{1}{(k_2)^2} (i_{k_1} i_{k_2} B) (i_{k_2} C^{(2)}) \wedge (i_{k_2} g) 
- \frac{k_1 \cdot k_2}{(k_2)^2} (i_{k_2} C^{(2)}) \wedge K^{(1)}
\right\} \wedge K^{(3)} \wedge K^{(2)}
\, , \\
\wt{C}^{(5)}
\ &= \ 
- i_{k_1} i_{k_2} C^{(7)} 
+ (i_{k_1} C^{(5)}) \wedge K^{(1)} 
- (i_{k_1} i_{k_2} B) C^{(5)} 
\nn \\
\ & \ \ \ \ 
+ \frac{1}{(k_2)^2} (i_{k_1} i_{k_2} C^{(5)}) \wedge K^{(1)} \wedge (i_{k_2} g) 
+ \frac{1}{(k_2)^2} (i_{k_1} i_{k_2} B) (i_{k_2} C^{(5)}) \wedge (i_{k_2} g) 
- \frac{k_1 \cdot k_2}{(k_2)^2} (i_{k_2} C^{(5)}) \wedge K^{(1)}
\nn \\
\ & \ \ \ \ 
- \left[
i_{k_2} C^{(5)} 
- C^{(3)} \wedge K^{(1)} 
- \frac{1}{(k_2)^2} (i_{k_2} C^{(3)}) \wedge K^{(1)} \wedge (i_{k_2} g)
\right] \wedge K^{(3)}
\nn \\
\ & \ \ \ \ 
+ \frac{(k_2)^2}{\det h_{IJ}}
\left[
i_{k_1} i_{k_2} C^{(5)} 
- (i_{k_1} C^{(3)}) \wedge K^{(1)} + (i_{k_1} i_{k_2} B) C^{(3)} 
- \frac{1}{(k_2)^2} (i_{k_1} i_{k_2} C^{(3)}) \wedge K^{(1)} \wedge (i_{k_2} g) 
\right. 
\nn \\
\ & \LS \LS
\left. 
- \frac{1}{(k_2)^2} (i_{k_1} i_{k_2} B) (i_{k_2} C^{(3)}) \wedge (i_{k_2} g) 
+ \frac{k_1 \cdot k_2}{(k_2)^2} (i_{k_2} C^{(3)}) \wedge K^{(1)}
\right] 
\wedge K^{(3)} \wedge K^{(1)}
\, .
\end{align}
\esubeq
Note that these transformation rules hold true only in the pull-back.

Finally, we exhibit the NS-NS $B$-field and the R-R fields that appear
in the effective action of the type IIB $5^2_3$-brane.
They are obtained by the S-duality transformations of the fields
$\wt{B}$, $\wt{C}^{(2)}$, $\wt{C}^{(4)}$.
Again in the pull-back, the explicit forms are found to be  
\bsubeq \label{eq:523backgrounds}
\begin{align} 
\wt{B}^{\prime} \ =& \ 
- C^{(2)} + \frac{1}{(k_2)^2} (i_{k_2} g) \wedge L^{(1)} + 
\frac{(k_2)^2}{\det l_{IJ}} L^{(2)} \wedge L^{(3)} 
\, , \\
\wt{C}^{(2)\prime} 
\ =& \ 
- i_{k_1} i_{k_2} (C^{(4)} + C^{(2)} \wedge B) 
+ (i_{k_1} B) \wedge L^{(1)} + (i_{k_1} i_{k_2} C^{(2)}) B 
- \frac{1}{(k_2)^2} (i_{k_1} i_{k_2} B) L^{(1)} \wedge (i_{k_2} g)
\nn \\
& \ 
+ \frac{1}{(k_2)^2} (i_{k_1} i_{k_2} C^{(2)}) (i_{k_2} B) \wedge
 (i_{k_2} g)
- \frac{k_1 \cdot k_2}{(k_2)^2} (i_{k_2} B) \wedge L^{(1)}
- (i_{k_2} B - C^{(0)} L^{(1)}) \wedge L^{(3)} 
\nn  \\
& \
- \frac{(k_2)^2}{\det l_{IJ}} 
\left(
i_{k_1} i_{k_2} (B - C^{(0)} C^{(2)}) L^{(3)} \wedge L^{(2)}
\right) 
\, , \\
\wt{C}^{(4) \prime} 
\ =& \ 
i_{k_1} i_{k_2} 
\left(
B^{(6)} - \frac{1}{2} B \wedge C^{(2)} \wedge C^{(2)}
\right) 
+ i_{k_1} 
\left(
C^{(4)} + C^{(2)} \wedge B
\right) \wedge L^{(1)} 
+ (i_{k_1} i_{k_2} C^{(2)}) (C^{(4)} + C^{(2)} \wedge B)
\notag \\
& \ 
- \frac{1}{(k_2)^2} i_{k_1} i_{k_2} (C^{(4)} + C^{(2)} \wedge B)
 \wedge L^{(1)} \wedge i_{k_2} g 
+ \frac{1}{(k_2)^2} (i_{k_1} i_{k_2} C^{(2)}) i_{k_2} (C^{(4)} + C^{(2)}
 \wedge B) \wedge i_{k_2} g 
\nn \\
& \
- \frac{k_1 \cdot k_2}{(k_2)^2} i_{k_2} (C^{(4)} + C^{(2)} \wedge B)
 \wedge L^{(1)}
\nn \\
& \ 
- \left[
i_{k_2} (C^{(4)} + C^{(2)} \wedge B) - B \wedge L^{(1)} 
+ \frac{1}{(k_2)^2} (i_{k_2} B) \wedge L^{(1)} \wedge i_{k_2} g 
\right] \wedge L^{(3)}
\nn \\
& \
- \frac{(k_2)^2}{\det l_{IJ}} 
\left[
i_{k_1} i_{k_2} 
(C^{(4)} + C^{(2)} \wedge B) 
- (i_{k_1} B) \wedge L^{(1)} 
- (i_{k_1} i_{k_2} C^{(2)}) B
+ \frac{1}{(k_2)^2} (i_{k_1} i_{k_2} B) L^{(1)} \wedge i_{k_2} g 
\right.
\nn \\
& \
\left.
- \frac{1}{(k_2)^2} (i_{k_1} i_{k_2} C^{(2)}) B \wedge i_{k_2} g 
+ \frac{k_1 \cdot k_2}{(k_2)^2} (i_{k_2} B) \wedge L^{(1)}
\right] \wedge L^{(3)} \wedge L^{(2)}
\, .
\end{align}
\esubeq

\end{appendix}

}
\end{document}